\documentclass[]{aa} 
\usepackage{epsfig}
\usepackage{wasysym}
\usepackage{graphicx}
\usepackage{natbib}
\usepackage{booktabs}
\usepackage{lscape}

\usepackage{url}
\usepackage[breaklinks]{hyperref}

\usepackage{color}
\providecommand{\tabularnewline}{\\}
\usepackage{txfonts}
  \def \teff {$T_{\mathrm{eff}}$}
  \def \tc {$T_{\mathrm{c}}$}
  \def \vtur {$V_{\mathrm{tur}}$}
  \def \zettwo {$\zeta^2$ Ret}
  \def \zetone {$\zeta^1$ Ret}

\begin{document}

  \title{$\zeta^2$ Ret, its debris disk, and its lonely stellar companion $\zeta^1$ Ret}
  \subtitle{Different \tc \ trends for different spectra\thanks{
  Based on data obtained from the ESO Science Archive Facility under request number vadibekyan204818, vadibekyan204820, and vadibekyan185979.}, 
  \thanks{The Tables with EWs of the lines and chemical abundances are only available in electronic form at the CDS via anonymous ftp to cdsarc.u-strasbg.fr (130.79.128.5)
or via http://cdsweb.u-strasbg.fr/cgi-bin/qcat?J/A+A/}}

  \author{V. Adibekyan\inst{1}  
          \and E. Delgado-Mena\inst{1}
          \and P. Figueira\inst{1}
          \and S. G. Sousa\inst{1}
          \and N. C. Santos\inst{1,2}
          \and J.~P.~Faria\inst{1,2}
          \and \\ J.~I.~Gonz\'{a}lez Hern\'{a}ndez\inst{3,4}
          \and G. Israelian\inst{3,4}
          \and G. Harutyunyan\inst{5} 
          \and L. Su\'{a}rez-Andr\'{e}s\inst{3,4}
          \and A. A. Hakobyan\inst{6}
         }

  \institute{
          Instituto de Astrof\'isica e Ci\^encias do Espa\c{c}o, Universidade do Porto, CAUP, Rua das Estrelas, 4150-762 Porto, Portugal\\
          \email{Vardan.Adibekyan@astro.up.pt}
          \and
          Departamento de F\'isica e Astronomia, Faculdade de Ci\^encias, Universidade do Porto, Rua do Campo Alegre, 4169-007 Porto, Portugal
          \and 
          Instituto de Astrof\'{\i}sica de Canarias, 38200 La Laguna, Tenerife, Spain
          \and 
          Departamento de Astrof{\'\i}sica, Universidad de La Laguna, 38206 La Laguna, Tenerife, Spain        
          \and 
          Leibniz Institute for Astrophysics Potsdam (AIP), An der Sternwarte 16, 14482 Potsdam, Germany
          \and
          Byurakan Astrophysical Observatory, 0213 Byurakan, Aragatsotn province, Armenia
}

  \date{Received date / Accepted date }
 
  \abstract
  {Several studies have reported a correlation between the chemical abundances of stars and condensation temperature (known as \tc \ trend). 
  Very recently, a strong \tc \ trend was reported for the $\zeta$ Reticuli binary system, which consists of two solar analogs. 
  The observed trend in \zettwo \ relative to its companion was explained by the presence of a debris disk around \zettwo.}
  {Our goal is to re-evaluate the presence and variability of the \tc \ trend in the $\zeta$ Reticuli system and to
  understand the impact of the presence of the debris disk on a star.}
  {We used very high-quality spectra of the two stars retrieved from the HARPS archive to derive very precise stellar parameters and chemical abundances. 
  We derived the stellar parameters with the classical (nondifferential) method, while we applied a differential line-by-line analysis  to 
  achieve the highest possible precision in abundances, which are fundamental to explore for very tiny differences in the abundances between the stars.}
  {We confirm that the abundance difference between \zettwo \ and \zetone \
  shows a significant ($\sim$ 2 $\sigma$) correlation with \tc. However, we also find that the \tc \ trends depend
  on the individual spectrum used (even if always of very high quality). In particular, we find significant but varying differences in 
  the abundances of the same star from different individual high-quality spectra.}
  {Our results for the $\zeta$ Reticuli system show, for example, that nonphysical factors, such as the quality of spectra employed and errors that are not accounted
for, can be at the root of the 
   \tc \ trends for the case of individual spectra.}
  

\maketitle

%

\section{Introduction}                                  \label{sec:intro}

In the last decade, noticeable advances were made in the characterization of atmospheric properties and chemical compositions of cool, sun-like stars. 
The achieved high precision in chemical abundances allowed observers to study  the chemical abundances of stars with and without planets in detail
\citep[e.g.,][]{Gilli-06, Robinson-06, Delgado-14, Delgado-15, Kang-11, Adibekyan-12a, Adibekyan-12b, Adibekyan-15a}. These studies are very important 
since the chemical abundances of stars with planets provide a huge wealth of information about the planet formation process 
\citep[e.g.,][]{Bond-10, Delgado-10}, composition of planets \citep[e.g.,][]{Thiabaud-14, Dorn-15, Santos-15}, and even about the habitability of the 
planets \citep[e.g.,][]{Adibekyan-16}.

The achieved extremely high precision in chemical abundances also facilitated the study of very subtle chemical peculiarities in stars that had initially appeared to be Sun-like. 
The quintessential example,  the so-called \tc \ trend, is the trend of individual chemical abundances with the condensation  temperature (\tc) of the elements. 
Many studies, beginning with \citet{Gonzalez-97} and \citet{Smith-01}, attempted to search for planet accretion and/or formation signatures in the photospheric 
chemical compositions of cool stars by looking at the \tc\ trends \citep[e.g.,][]{Takeda-01, Ecuvillon-06, Sozzetti-06, Melendez-09, Ramirez-09, GH-10, 
Schuler-11, GH-13, Maldonado-15, Nissen-15, Biazzo-15, Saffe-15}.

\citet{Melendez-09} claimed that the Sun shows a deficiency in refractory elements with respect to other solar twins because 
these elements were trapped in the terrestrial planets in our solar system. The same conclusion was also reached by \citet{Ramirez-09}, who analyzed 64 solar twins and
analogs with and without detected planets. However, later results by \citet{GH-10} and \citet{GH-13} strongly challenge the relation between the presence of 
planets and the abundance peculiarities of the stars.

Rocky material accretion is by far not the only explanation for the \tc  \ trend. Recently, \citet{Adibekyan-14} showed that the \tc\ trend strongly
depends on the stellar age\footnote{We note that \citet{Ramirez-14} also observed a correlation between the \tc   \ slope and stellar age
for metal-rich solar analogs, but apparently with opposite sign. In contrast to our results, they found that
most refractory element depleted stars are younger than the those with the highest refractory element abundances.} %
and they found a tentative dependence on the galactocentric distances of the stars. The authors concluded that the \tc\ trend depends on the Galactic
chemical evolution (GCE) and suggested that the difference in  \tc \ slopes, observed between planet hosting solar analogs and solar analogs without detected 
planets, may reflect their age difference. \citet{Maldonado-15} and  \citet{Maldonado-16} also confirmed the correlation of the \tc\ trend with age, and further suggested a 
significant correlation with the stellar radius and mass. \citet{Onehag-14} in turn showed that while the Sun shows a different \tc\ trend when compared to the 
solar-field twins, it shows a very similar abundance trend with \tc \ when compared to the stars from the open cluster M67. These authors suggested that the Sun, 
unlike most stars, was formed in a dense stellar environment where the protostellar disk was already depleted in refractory elements before the star formation. 
\citet{Gaidos-15} also suggested that gas-dust segregation in the disk can be responsible for the \tc\ trends.

While above mentioned mechanisms and processes (e.g., GCE or age effect) can be responsible for the general \tc\ trends for field-star samples, 
 they cannot easily explain trends observed between companions of binary systems. Several authors studied the \tc\ trend in binary stars with and without planetary
companions \citep[e.g.,][]{Liu-14, Saffe-15, Mack-16} or binary stars where both components host planets \citep[e.g.,][]{Biazzo-15, Teske-15, Ramirez-15, Teske-16}. 
The results and conclusions of these studies show that there are no systematic differences in the chemical abundances of 
stars with and without planets in binary systems. Moreover, there is no consensus on the results even for the same, individual systems \citep[e.g., 16 Cyg AB;][]{Laws-01, Takeda-05,
Schuler-11a, TucciMaia-14}.

Recently, \citet[][]{Maldonado-15} studied a large sample of stars with and without debris disks to search for chemical anomalies related to the formation of planets.
They found no significant differences in chemical abundances or in the \tc\ trends between the two samples. However, very recently, \citet[][ hereafter S16]{Saffe-16}
reported a positive \tc \ trend in the binary system, \zetone \ -- \zettwo, \ where one of the stars (\zettwo) hosts a debris disk. 
The authors explained the deficit of the refractory elements relative to volatiles in \zettwo\  as caused by the depletion of about $\sim$3 M$_{\oplus}$  rocky material.

The two stars have very similar atmospheric parameters, which in principle allows for a very high-precision relative abundance characterization of the stars.
Taking full advantage of this, S16 carried out a detailed and careful analysis of the system. However, the authors chose to use only the spectra
of the stars observed during the same night with the same instrument (High Accuracy Radial velocity Planet Searcher -- HARPS). The final signal-to-noise ratio (S/N) of the spectra that they used was $\sim$ 300
(S16), while we noticed that much higher S/N (see Table\,\ref{tab:sample}) can be achieved 
if all of the spectra from the ESO (European Southern Observatory) HARPS archive\footnote{http://archive.eso.org/wdb/wdb/adp/phase3\_spectral/form} are combined. Moreover, some of the individual 
spectra found in the archive have S/N of more than 350, which allows us to carry out a differential abundance
analysis between different spectra of the same star, which in turn provides an independent estimation of the precision of our measurements.

This case motivated us to explore whether the observed differences in the chemical abundances of the stars could be simply due to some
systematics in the spectra.
In this work, we use very high-quality spectra of the \zetone \ and \zettwo \ binary star system to re-evaluate the presence and variability of the \tc \ 
trend in this system and, as such, better understand the impact of the presence of the debris disk on a star. We organized this paper
as follows. In Sect.\,\ref{sec:data} we  present the data, in Sect.\,\ref{sec:parameters}
we present the methodology used to derive the stellar parameters and chemical abundances, and in Sect.\,\ref{sec:tc_slope} we explain how we 
calculate and evaluate the significance of the \tc \ trends. After presenting the main results in Sect.\,\ref{sec:results}, we  
conclude in Sect.\,\ref{sec:conclusion}.

\begin{table*}[t!]

\setlength{\tabcolsep}{3pt}
\caption{\label{tab:sample} Stellar parameters, S/N, and observation dates of \zetone \ and \zettwo \ derived from individual and combined spectra.}
\centering
\begin{tabular}{lcccccc}
\hline\hline
Star(spectrum) & \teff \ & $\log g$ & [Fe/H] & \vtur & S/N  & DATE-OBS\\
\hline
sun\_vesta & 5777$\pm$10 & 4.43$\pm$0.02 & 0.020$\pm$0.010 & 0.95$\pm$0.02 & 1340  & -- \tabularnewline
\zettwo\_a & 5856$\pm$15 & 4.49$\pm$0.02 & -0.205$\pm$0.012 & 0.95$\pm$0.02 & 470 & 2006-11-08T04:02:51.439 \tabularnewline
\zettwo\_b & 5837$\pm$12 & 4.47$\pm$0.02 & -0.215$\pm$0.010 & 0.96$\pm$0.02 & 450 & 2009-08-27T08:42:10.196 \tabularnewline
\zettwo\_c & 5835$\pm$13 & 4.48$\pm$0.03 & -0.216$\pm$0.011 & 0.94$\pm$0.02 & 420 & 2010-01-31T01:32:05.689 \tabularnewline
\zettwo\_comb & 5861$\pm$12 & 4.53$\pm$0.02 & -0.215$\pm$0.010 & 1.00$\pm$0.02 & 3000 & -- \tabularnewline
\zettwo\_S16 & 5833$\pm$15 & 4.48$\pm$0.03 & -0.218$\pm$0.012 & 0.95$\pm$0.03 & 230 & 2004-02-04 \tabularnewline
\zetone\_a & 5713$\pm$14 & 4.48$\pm$0.02 & -0.197$\pm$0.011 & 0.88$\pm$0.02 & 375 & 2005-11-15T03:41:34.591 \tabularnewline
\zetone\_b & 5696$\pm$14 & 4.47$\pm$0.02 & -0.197$\pm$0.011 & 0.81$\pm$0.02 & 370 & 2005-11-15T03:52:37.899 \tabularnewline
\zetone\_c & 5702$\pm$13 & 4.47$\pm$0.03 & -0.203$\pm$0.011 & 0.85$\pm$0.03 & 360 & 2005-11-15T03:47:07.475 \tabularnewline
\zetone\_comb & 5720$\pm$13 & 4.52$\pm$0.03 & -0.206$\pm$0.010 & 0.89$\pm$0.02 & 1300 & -- \tabularnewline
\zetone\_S16 & 5722$\pm$15 & 4.50$\pm$0.03 & -0.191$\pm$0.011 & 0.86$\pm$0.02 & 215 & 2004-02-04 \tabularnewline
\hline
\end{tabular}
\end{table*}

\section{Data}                                          \label{sec:data}

Both stars of this binary system were extensively observed with the HARPS high-resolution spectrograph \citep{Mayor-03} at the 3.6 m telescope (La Silla Paranal Observatory, ESO).
The HARPS archive contains $\sim$70 and $\sim$170 spectra for \zetone \ and \zettwo, respectively.  We use the combined spectra
of the stars (referred as 'starname\_comb' in Table\,\ref{tab:sample}), spectra that were used in S16 (referred as 'starname\_S16' in Table\,\ref{tab:sample}),
and three individual spectra with the highest S/N for each star (referred as 'starname\_a', 'starname\_b', and 'starname\_c' in Table\,\ref{tab:sample}).
The highest S/N spectra are used to provide the final characteristics of this system. The individual high S/N spectra are selected to estimate the internal
precision of the methods, while the spectra used by S16 are selected for external comparison. 

A few  of the spectra of these stars were obtained under poor atmospheric conditions (the headers of the fits files report -1.00 value for seeing) ,
and we considered these spectra unsuitable for this work and, therefore, chose not to use them. These 
spectra had very low S/N and  combining them would increase the total S/N of the combined spectra (listed in Table\,\ref{tab:sample}) 
by 0.03 and 1.6\% for \zetone \ and \zettwo, respectively.
The S/N at the 40th fiber order (central wavelength $\sim$ 5060\AA{}) for each of the HARPS spectra (retrieved from the header of the spectra) are presented in Table\,\ref{tab:sample}\footnote{
The S/N for the combined spectra are calculated as a quadrature sum of the individual S/Ns.}. As one can see, even the individual spectra of the stars 
have a much higher S/N than that of the spectra used in S16 and, in the combined spectra, the S/N are higher by a factor of about six and $\sim$13 for \zetone \ and \zettwo, \ respectively. 
We note again that S16 chose to use the spectra of the stars obtained during the same night, probably to minimize the time-dependent
systematics.
For the Sun, we used a combined HARPS reflected spectrum from Vesta (extracted from the same public archive, S/N $\sim$1300).

\section{Stellar parameters and chemical abundances}            \label{sec:parameters}

\begin{figure}
\begin{center}
\begin{tabular}{c}
\includegraphics[angle=0,width=1.0\linewidth]{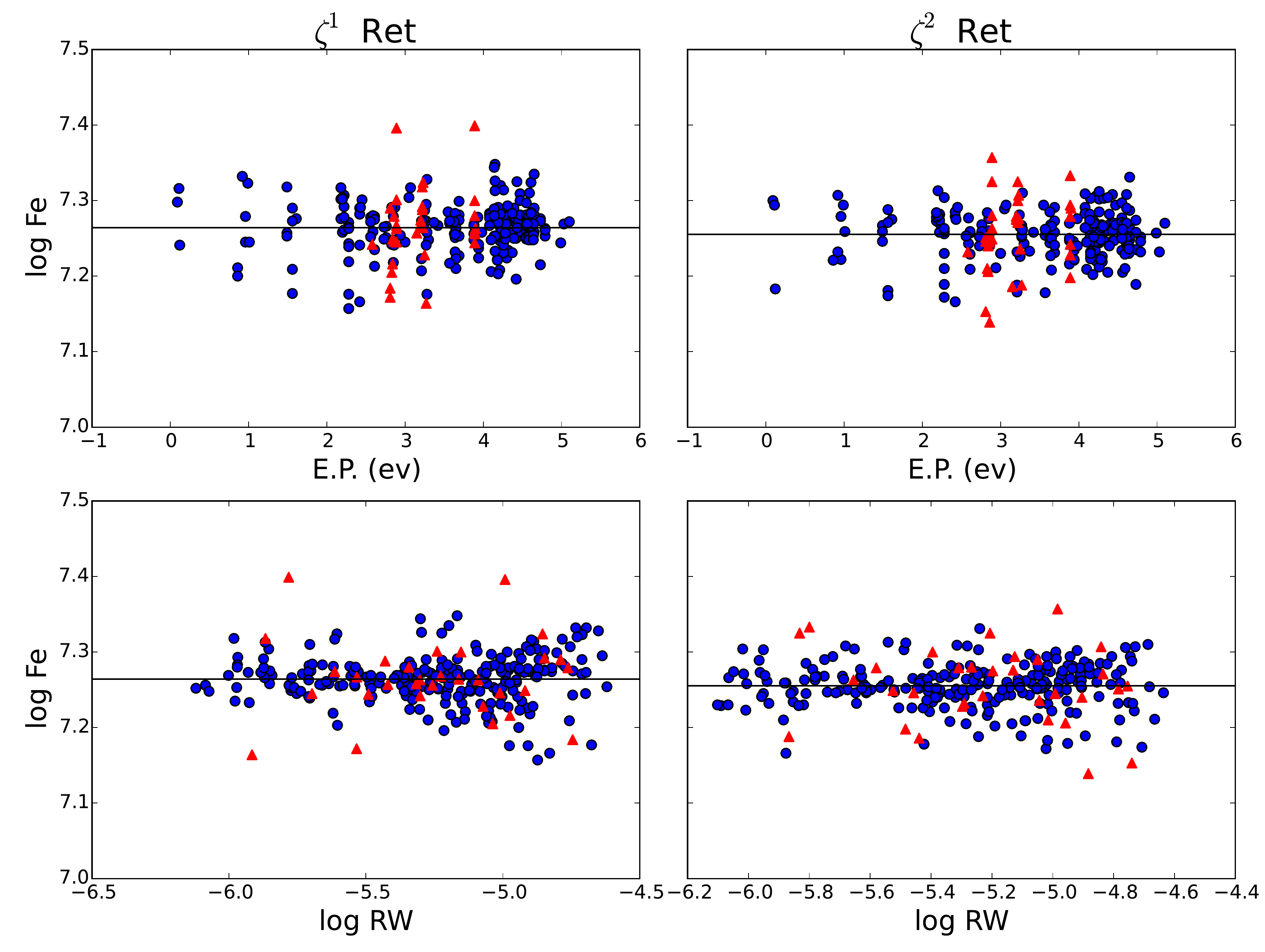}
\end{tabular}
\end{center}
\vspace{-0.5cm}
\caption{Iron abundance vs. excitation potential (upper panels) and iron abundance vs. reduced EW (lower panels) for \zetone\_comb and \zettwo\_comb.
The blue and red symbols correspond to the neutral and ionized iron lines. The black solid line indicates the linear fit of the data.}
\label{plot_fe_param}
\end{figure}

We derived the stellar parameters (\teff, [Fe/H], $\log g$, and \vtur) for the stars from the various spectra and for the Sun  with the procedure described in 
\citet[][]{Sousa-14}. In short, we first automatically measured the equivalent widths (EWs) of iron lines ($\sim$250 \ion{Fe}{i} and $\sim$40 \ion{Fe}{ii} lines) 
using the ARES v2 code\footnote{The last version of ARES code (ARES v2) can be downloaded at http://www.astro.up.pt/$\sim$sousasag/ares} \citep{Sousa-15}.
Then the spectroscopic parameters were derived by imposing excitation and ionization balance assuming LTE (see Fig.~\ref{plot_fe_param}). 
We used the grid of ATLAS9 plane-parallel model
of atmospheres \citep{Kurucz-93} and the 2014 version of MOOG\footnote{The source code of MOOG can be downloaded at
http://www.as.utexas.edu/$\sim$chris/moog.html}  radiative transfer code \citep{Sneden-73}. 
We derived the parameters of the stars  with a classical rather than a line-by-line differential approach.
In general, when stars with similar stellar parameters are compared, the two methods give very similar results and the estimated errors in case of differential
line-by-line analysis are slightly smaller. The uncertainties of the parameters are derived as in our previous works 
and they are well described in \citet[][]{Neuforge-97}. The derived stellar parameters are listed in Table\,\ref{tab:sample}. 

The values in Table\,\ref{tab:sample} show that the stellar parameters of the same stars derived from different spectra agree very well within the 
estimated errors. Furthermore, when the same spectra are used, our atmospheric parameters agree very well with those derived in S16.
In particular, we obtained $\Delta$\teff \ = 12 K, $\Delta$[Fe/H] = 0.004 dex, $\Delta\log g$ = -0.03 dex, 
and $\Delta$\vtur \ = 0.06 km s$^{-1}$ for \zetone, \ where the differences are defined as our values minus those of S16.
For \zettwo, \, the differences are $\Delta$\teff \ = -21 K, $\Delta$[Fe/H] = -0.003 dex, $\Delta\log g$ = -0.06 dex, and 
$\Delta$\vtur \ = 0.00 km s$^{-1}$. 

\begin{figure}
\begin{center}
\begin{tabular}{c}
\includegraphics[angle=0,width=1.0\linewidth]{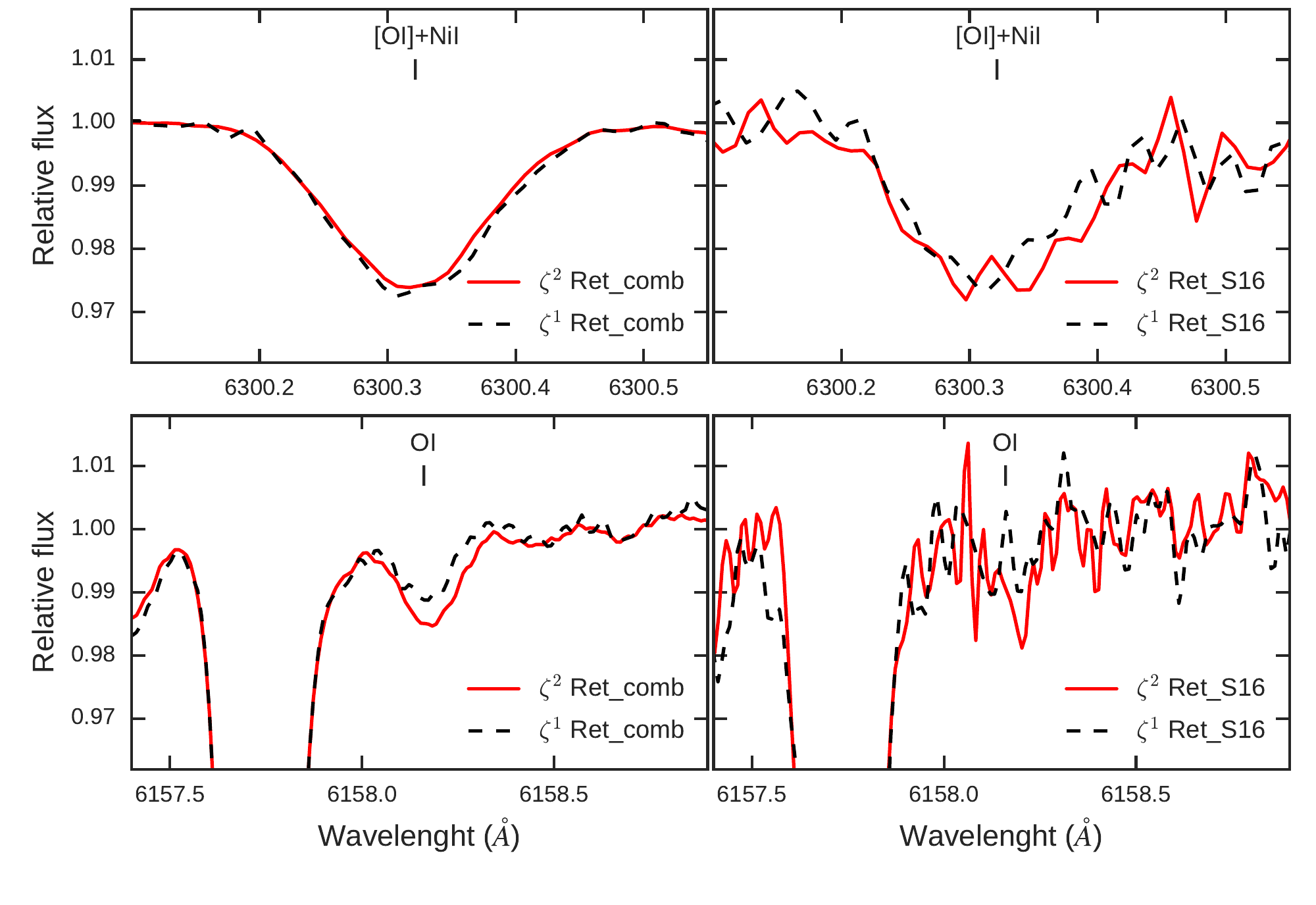}
\end{tabular}
\end{center}
\vspace{-0.5cm}
\caption{Spectral regions containing oxygen lines in \zetone\_comb, \zettwo\_comb, \zetone\_S16, and \zettwo\_S16 spectra.}
\label{plot_oxygen}
\end{figure}

\begin{figure*}
\begin{center}
\begin{tabular}{c}
\includegraphics[angle=0,width=1.0\linewidth]{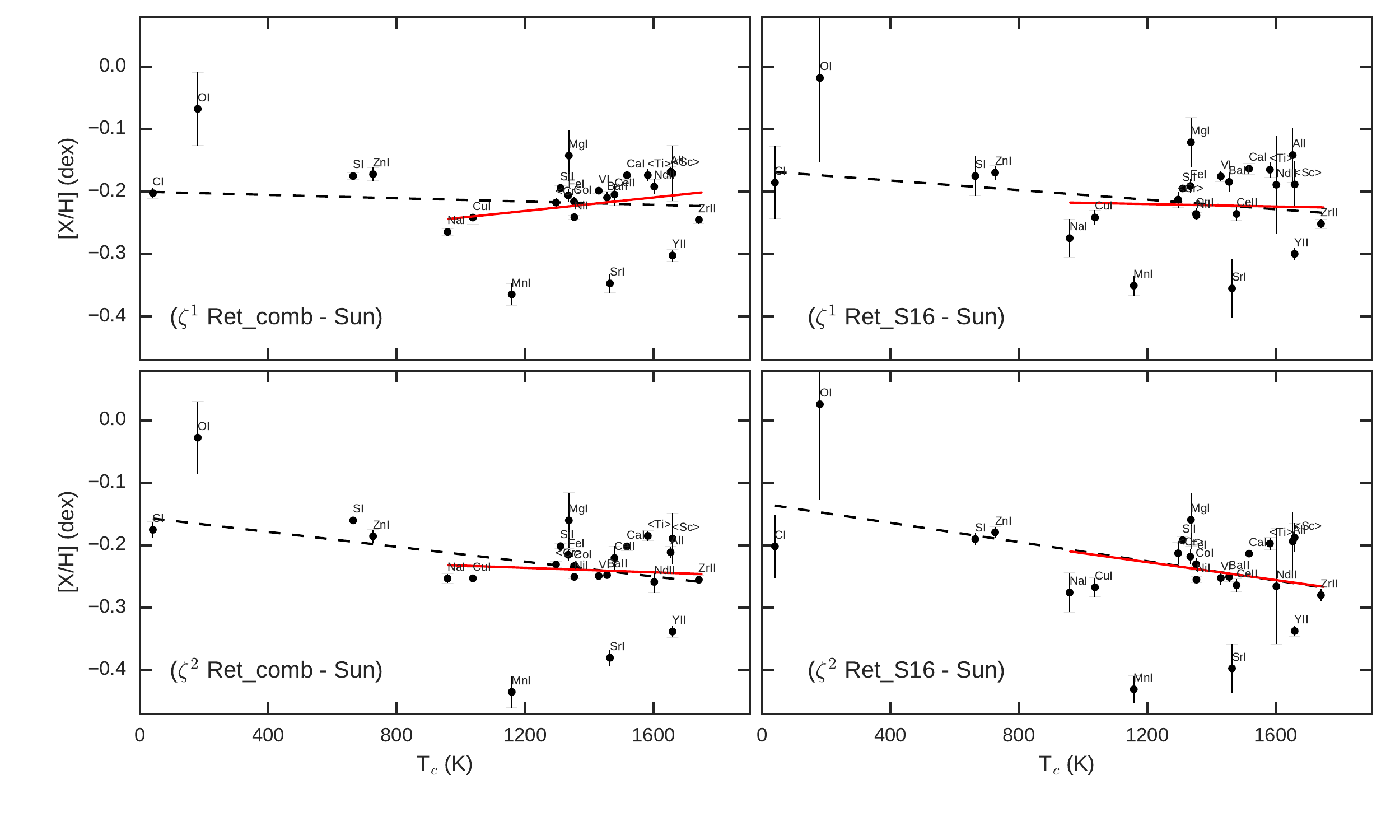}
\end{tabular}
\end{center}
\vspace{-1.0cm}
\caption{Differential abundances [X/H] against condensation temperature  for \zetone \ and \zettwo . The abundances  are derived relative to the Sun with the combined spectra and spectra from S16. The black dashed line represents the trend when all of the elements are used for the linear regression and 
the red line is the result of the linear regression when only elements with \tc \ $>$ 900K are used.}
\label{fig_rel_sun}
\end{figure*}

Similar to S16, we derived 0.027 dex higher metallicity for \zetone \ compared to \zettwo \ with the same spectra. However, the difference
decreases to 0.009 dex when the spectra with the highest S/N ratio are considered. Also, when  different pairs of spectra of the stars are compared, 
the maximum difference is obtained when the \zetone\_S16 and \zettwo\_S16 are used, and when \zetone\_comb and \zettwo\_a pairs are compared, the \zettwo \ 
appears even slightly more metal rich than \zetone \ by -0.001 dex. These results suggest that one should be cautious when 
reaching such an extremely small difference in stellar parameters.

Elemental abundances for the stars were also determined using an LTE analysis and the same tools and codes as for stellar parameters determination.
We adopted the initial line list for Na, Mg, Al, Si, Ca, Ti, Cr, Ni, Co, Sc, Mn, and V from \citet{Adibekyan-12c},
but several lines (two \ion{Si}{i} lines $\lambda$5701.11, $\lambda$6244.48, two \ion{Ca}{i} lines $\lambda$5261.71, $\lambda$5352.05, 
three \ion{Ti}{i} lines $\lambda$4722.61, $\lambda$5039.96, $\lambda$5965.84, one \ion{Ti}{ii} line $\lambda$ 5381.03,
two \ion{Cr}{i} lines $\lambda$5122.12, $\lambda$6882.52, and two \ion{Ni}{i} lines $\lambda$5081.11, $\lambda$6767.78) 
were excluded because of large [X/Fe] star-to-star scatter at solar metallicities \citep{Adibekyan-15b}. Finally, \ion{Ni}{i} line at  $\lambda$6360.81
was excluded because of unreliable EW measurements (because of a depressed continuum) and the \ion{Sc}{i} line at  $\lambda$5520.50 was excluded because 
there is no data about the hyperfine structure splitting (HFS) of the lines in
Kurucz atomic line database\footnote{http://kurucz.harvard.edu/linelists.html} nor in \citet{Prochaska-00}, which are our two sources for 
the relative $\log gf$  values.

We determined oxygen abundances using two weak lines at 6158.2\AA{} and 6300.3\AA{} following the work of 
\citet{Bertrandelis-15}, although it was only possible to measure the 6158.2\AA{} lines for the combined spectra of \zetone \ and \zettwo\  and 
for the three individual spectra of \zettwo.
Sulfur abundances were calculated by performing spectral synthesis with MOOG around 
the lines 6046.0\AA{}, 6052.5\AA{}, 6743.5\AA,{} and 6757.1\AA{}. The atomic data of those lines and the 
surrounding lines were taken from VALD3 database\footnote{http://vald.astro.univie.ac.at/~vald3/php/vald.php?newsitem=0}. 
The lines at 6046.0\AA{} and 6052.5\AA{} (used in S16) suffer from non-negligible blends of CN bands, thus, our initial 
abundances derived from EWs yielded higher abundances than the other two lines.
To fit the lines in the solar spectra, we had to calibrate the $\log gf$s of the S lines (the line list and atomic data  are available at the CDS), to match a solar sulfur abundance A(S)=7.16 \citep{Asplund-09}. 
Cerium abundances were derived using the three spectral lines presented in \citet{Reddy-03} and the line at 5274.2\AA{} for which the 
atomic data was extracted from VALD3.
Carbon abundances were based on the two well-known CI optical lines at 5052\AA{} and 5380\AA{}. 
The atomic data for carbon and elements heavier than Ni were also extracted from VALD3 
(a more detailed line list will be shown in Delgado Mena et al. in prep). 

When available, the Barklem damping van der Waals constants were used
for all of the elements.
Otherwise, the Uns\"{o}ld  approximation multiplied  by a factor (1.0+0.67$\times$E.P., where E.P. is the excitation potential of a line) suggested by the Blackwell 
group was used (option 2 in the damping parameter inside MOOG).
For the Sc, V, Mn, Co, Cu, and Ba lines,  HFS was considered.  We adopted the atomic parameters and isotopic ratios from \citet{Prochaska-00} for Ba, and the relative $\log gf$ values and isotopic ratios for the lines were taken from Kurucz database for
the remaining elements.

The EWs of the lines were derived with ARES v2 code, but with careful visual inspection. In a few cases, when the ARES measurements were not 
satisfactory (this can happen, e.g., as a result of a presence of cosmic rays or bad pixels), we measured the EWs using the task \emph{splot} 
in IRAF\footnote{IRAF is distributed by National Optical Astronomy Observatories,
operated by the Association of Universities for Research in Astronomy, Inc., under contract with the National Science Foundation, USA.}. 
We calculated the final abundance of the elements (when several spectral lines were available) as a weighted mean of all
of the abundances, where the distance from the median abundance was considered  as a weight. As demonstrated in \citet{Adibekyan-15b}, this method can be effectively
used without removing suspected outlier lines.

 We performed differential line-by-line analysis relative to the Sun for the combined and S16 spectra. We also performed differential
abundance analysis of \zettwo\_comb  relative to \zetone\_comb, and \zettwo\_S16 relative to \zetone\_S16.
Finally, for both stars we derived differential abundances from each of the individual three spectra 
('starname\_b' and 'starname\_c' relative to 'starname\_a') relative to each other.

\begin{table*}[t!]

\setlength{\tabcolsep}{3pt}
\caption{\label{tab:slopes} Slopes of the [X/H] vs. \tc \ for different pairs of spectra. Both frequentist and Bayesian approaches are chosen to derive 
the slopes and their uncertainties.}
\tiny
\begin{center}
\begin{tabular}{lccccc|ccccc}

\hline\hline
Star -- Reference &\multicolumn{3}{c}{WLS (all elements)} & \multicolumn{2}{c}{Bayesian (all elements)} & \multicolumn{3}{c}{WLS (refractories)} & \multicolumn{2}{c}{Bayesian (refractories)}\\ \cmidrule(l){2-4}\cmidrule(l){5-6}\cmidrule(l){7-9}\cmidrule(l){10-11}
 & \multicolumn{1}{c}{Slope$\pm \sigma$} & 95\% CI &  P(F-stat) & Slope & \multicolumn{1}{c}{95\% HPD} & \multicolumn{1}{c}{Slope$\pm \sigma$} & 95\% CI &  P(F-stat) & Slope & \multicolumn{1}{c}{95\% HPD}\\
\hline
\zetone\_comb -- Sun  & -1.32$\pm$2.28 & [-6.04, 3.40] & 0.567 & -1.32 & [-2.11, -0.46] & 5.40$\pm$ 4.37 & [-3.78, 14.58] & 0.232 & 5.37 & [3.75, 6.99] \tabularnewline
\zetone\_S16 -- Sun  & -3.83$\pm$3.92 & [-11.96, 4.30] & 0.339 & -3.83 & [-5.67, -1.75] & -0.99$\pm$ 5.84 & [-13.25, 11.28] & 0.868 & -1.01 & [-3.56, 1.75] \tabularnewline
\zettwo\_comb -- Sun  & -5.95$\pm$2.66 & [-11.46, -0.44] & 0.036 & -5.95 & [-6.93, -4.84] & -1.81$\pm$ 5.07 & [-12.45, 8.83] & 0.725 & -1.81 & [-3.78, -0.02] \tabularnewline
\zettwo\_S16 -- Sun  & -7.68$\pm$3.19 & [-14.30, -1.07] & 0.025 & -7.68 & [-9.22, -6.12] & -7.16$\pm$ 6.89 & [-21.63, 7.31] & 0.312 & -7.19 & [-10.22, -4.26] \tabularnewline
\zetone\_comb -- \zettwo\_comb  & 3.41$\pm$1.23 & [0.86, 5.96] & 0.011 & -3.41 & [-4.27, -2.52] & 4.88$\pm$ 2.16 & [0.35, 9.41] & 0.036 & 4.88 & [3.45, 6.29] \tabularnewline
\zetone\_S16 -- \zettwo\_S16  & 3.77$\pm$2.16 & [-0.71, 8.25] & 0.095 & -3.79 & [-5.37, -2.00] & 4.83$\pm$ 3.42 & [-2.34, 12.01] & 0.174 & 4.86 & [2.52, 7.31] \tabularnewline
\zetone\_b -- \zetone\_a  & 2.35$\pm$2.21 & [-2.24, 6.94] & 0.300 & 2.34 & [1.17, 3.62] & 4.83$\pm$ 2.83 & [-1.12, 10.77] & 0.105 & 4.83 & [3.34, 6.44] \tabularnewline
\zetone\_c -- \zetone\_a  & -0.14$\pm$0.84 & [-1.89, 1.61] & 0.867 & -0.14 & [-1.30, 0.99] & -0.22$\pm$ 1.40 & [-3.16, 2.73] & 0.880 & -0.21 & [-1.91, 1.59] \tabularnewline
\zetone\_c -- \zetone\_b  & 1.13$\pm$0.76 & [-0.45, 2.70] & 0.153 & 1.13 & [0.24, 2.00] & -0.28$\pm$ 0.90 & [-2.18, 1.61] & 0.759 & -0.28 & [-1.45, 0.80] \tabularnewline
\zettwo\_b -- \zettwo\_a  & -1.60$\pm$0.79 & [-3.24, 0.04] & 0.055 & -1.61 & [-3.09, -0.18] & -1.67$\pm$ 1.04 & [-3.84, 0.51] & 0.125 & -1.68 & [-3.43, 0.12] \tabularnewline
\zettwo\_c -- \zettwo\_a  & -1.36$\pm$0.76 & [-2.94, 0.22] & 0.087 & -1.35 & [-2.59, -0.17] & -0.35$\pm$ 0.94 & [-2.33, 1.63] & 0.714 & -0.37 & [-2.16, 1.32] \tabularnewline
\zettwo\_c -- \zettwo\_b  & -1.86$\pm$1.06 & [-4.05, 0.33] & 0.092 & -1.86 & [-3.24, -0.44] & 0.72$\pm$ 1.37 & [-2.15, 3.59] & 0.603 & 0.70 & [-1.52, 3.09] \tabularnewline
\hline
\end{tabular}
\end{center}
Note:$^{*}$ The units of the slopes are in 10$^{-5}$ dex K$^{-1}$.
\end{table*}

The errors of the [X/H] abundances are calculated as a quadrature sum of the errors due to EW measurements and errors due to uncertainties in the atmospheric
parameters. When three or more lines were available, the EW measurement error is estimated as $\sigma /\sqrt{(n-1)}$, where $\sigma$ is 
the line-to-line abundance scatter and $n$ is the number of the observed lines. The errors arising from uncertainties in the stellar parameters 
are calculated as a quadratic sum of the abundance sensitivities on the variation of the stellar parameters by their one $\sigma$ uncertainties.
 These errors are usually much smaller (because of very small uncertainties in stellar parameters) than the errors due to EW measurements.

 The error estimation method described in the previous paragraph is most frequently used, even when only two lines are observed. However,
the dispersion ($\sigma_{sample}$) calculated using only two lines can be very different from the dispersion rooted in the uncertainty on the abundances, 
a value that can only be efficiently estimated if large number of lines are available.
A simple calculation shows that if one assumes that the errors follow a normal distribution and  that $\sigma_{sample}$ is the standard deviation of that
distribution, then the $\sigma_{two-lines}$ calculated from  two random lines from that distribution peaks close to zero, underestimating $\sigma_{sample}$.
The $\sigma_{two-lines}$ is smaller than 0.5$\sigma_{real}$ in  $\sim$38\% cases and $\sigma_{two-lines}$ is five times 
smaller than $\sigma_{real}$ in $\sim$16\% cases. Moreover, in the cases when two lines of an element give exactly the same abundance, then the calculated dispersion is zero
and the final error is equal to the error due to uncertainties in the atmospheric parameters. Since the latter is usually very small, the weight given to
this element when calculating the \tc \ slope is extremely high (see Sect.\,\ref{sec:tc_slope}). Such underestimated errors can play a crucial role 
in the incorrect determination of the slopes.

We calculated the errors on EWs following \citet[][]{Cayrel-98} to provide more realistic errors for the abundances of elements that 
only have two observed spectral lines in our spectra (except oxygen) .  The calculated uncertainty takes into account the statistical photometric error due to the noise in each pixel and 
the error related to the continuum placement, which is the dominant contribution to the error \citep[][]{Cayrel-98, Bertrandelis-15}. Then, these  
errors are propagated to derive the abundance uncertainties for each line. The final uncertainties for the average abundance are propagated from these individual errors. 

When only one line for a given element is available, as is the case for O (for some spectra) and Sr, we determine the error by 
measuring a second EW with the position of the continuum displaced within the root mean square \textit{(rms)} 
due to the noise of the spectra, and  calculating the difference in abundance with respect to the original value. In this manner, the elements with only a weak line 
(as for oxygen in some cases) have a large error in their abundance measurements. This is in contrast to elements with strong lines, such as Sr, where the fact 
of having a single line does not affect the final error too much; see, for comparison, the error bars in Fig.~\ref{fig_rel_sun} for O and Sr and  how they vary 
between the highest and lowest S/N spectra). We stress here the case of oxygen for which we find a very different value than S16. 
Their differential abundance (\zetone \ -- \zettwo) is almost 0.1 dex lower than ours and their reported error is much smaller. 
In Fig.~\ref{plot_oxygen} we show a part of the spectra used by S16 around the 6300\AA{} region; it is strongly affected by noise. 
The EW values are very similar for both stars, but our abundance determinations have a larger error because of the strong 
effect of a small change in the continuum in this very weak line.
As mentioned by S16, this line is blended with a Ni line, and if we consider that the two stars have similar 
Ni abundances, it seems difficult 
to reconcile such a large difference in abundance values even if our atomic parameters for the Ni blend are different. However, for the combined, very high S/N spectra we can also
measure with better precision the oxygen line at 6158.2\AA,{} which suggests very similar oxygen abundance to that derived from the 
6300.3\AA{} line. This example shows the difficulty in measuring 
oxygen abundances. Such an issue is very important in the context of \tc \ trend analysis, as the low condensation 
temperature of oxygen leads to a strong leverage on the final slope value. 

Since both oxygen lines are very weak and deserve special attention \citep[see, e.g.,][]{Bertrandelis-15}, even when both lines were 
observed for a spectrum we calculated individual errors for each line (as described in the previous paragraph) and then propagated the error of the average 
oxygen abundance.
The EWs of all of the measured lines and the abundances of the elements derived from all of the spectra are available on the CDS.

\section{\tc \ slope}           \label{sec:tc_slope}

Once the differential abundances and corresponding errors are derived we can search for abundance trends with the condensation temperature of the elements. 
The 50\% \tc \ equilibrium condensation temperatures for a gas of solar system composition are taken from \citet{Lodders-03}. It is common practice to plot
[X/Fe] (and not just [X/H]) versus \tc \ , which allows us to remove the trends related to Galactic chemical evolution \citep[e.g.,][]{GH-13, Saffe-16}. Although
this procedure can be justified when stars with different metallicities are compared, it can produce a bias in the derived slope. 
Subtracting the iron abundance (a constant value) from each elemental
abundance should only shift the trend but should not change the slope. However, since iron abundance has an error itself, the propagated errors of the [X/Fe] abundance ratios
should be adjusted. This change in turn can have an impact on the original slope. In an extreme (although not very realistic) case that the error of iron abundance is much larger
than the errors of individual [X/H] abundances, then the propagated errors of [X/Fe] abundance ratios would be dominated by the error of [Fe/H]. In this case the 
weighted least-squares regression would give a result similar to an ordinary least-squares regression, i.e., the weights of all of the elements would be similar. This discussion
suggests that if the abundances are normalized to another element (not iron), the result would be different depending on the error of the element used
for normalization. Moreover, if the abundances are not normalized to the iron abundance, the [Fe/H] value can also be used in the calculations of the slopes. 
For these reasons we use [X/H] in our analysis when deriving the \tc \ trend.

In Table\,\ref{tab:slopes} we present the \tc  \ slopes, corresponding standard errors, 95\% confidence intervals (CI) and p-values (at $\alpha$=0.05 significance level). 
The \tc \ slopes refer to the slope of the linear dependence between [X/H] and \tc . We calculated the slopes with the weighted least-squares (WLS)
technique, whereas we calculated the weights as the inverse of the variance ($\sigma^2$) of the abundance. The p-values come from 
the F-statistics that tests the null hypothesis that the data can be modeled accurately by setting the regression coefficients to zero. The 95\% confidence intervals
are calculated using the standard error of the slopes and the p-value from T-statistics that tests the null hypothesis that the 
coefficient of a predictor variable is zero, i.e., slope is zero. 

\begin{figure}
\begin{center}
\begin{tabular}{c}
\includegraphics[angle=0,width=1\linewidth]{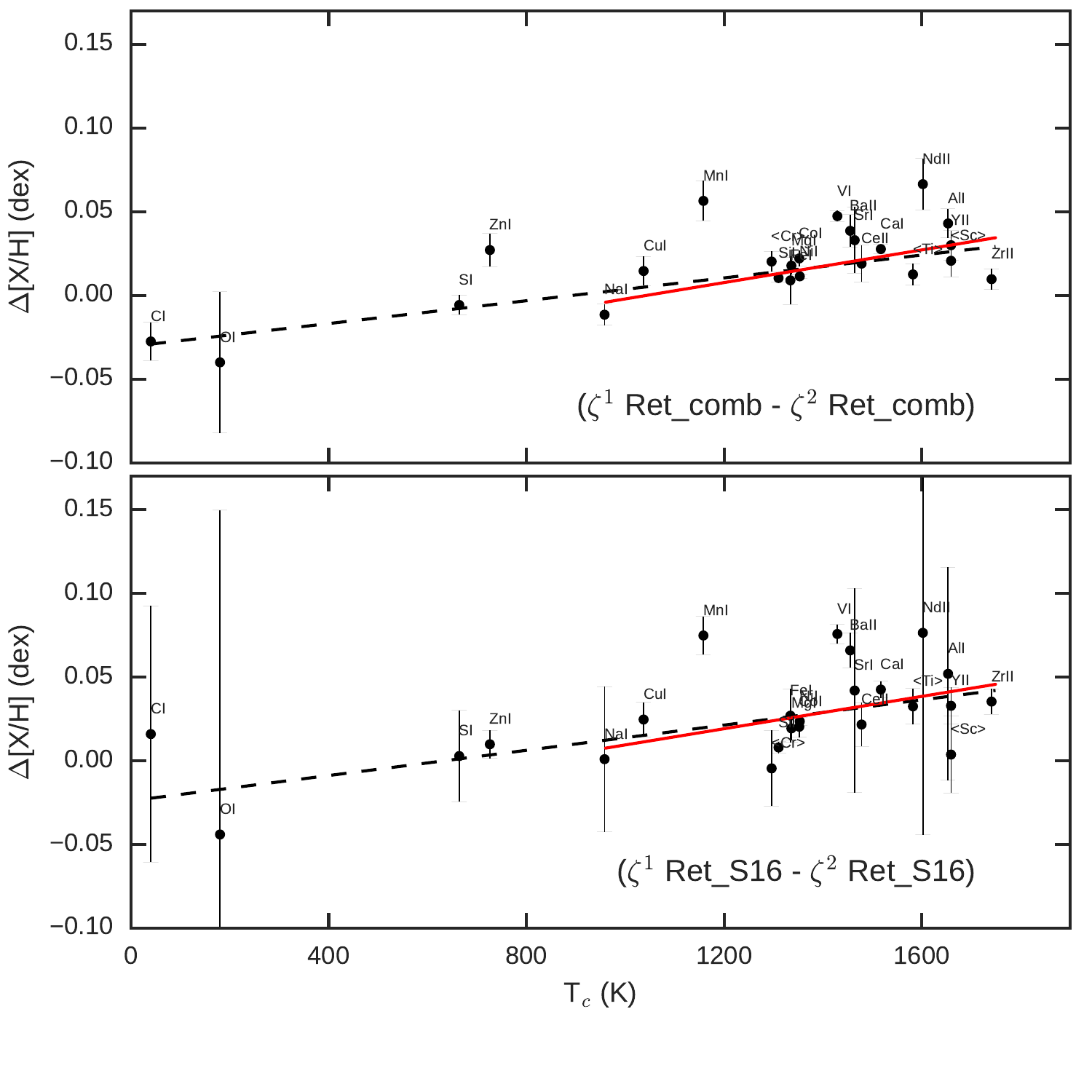}
\end{tabular}
\end{center}
\vspace{-1.cm}
\caption{Differential abundances (\zettwo \ -- \zetone ) against condensation temperature. The abundances  are derived for the combined spectra and spectra that were used in S16. The black dashed line and red line are the same as in Fig.~\ref{fig_rel_sun}.}
\label{fig_rel_z1}
\end{figure}

\begin{figure*}
\begin{center}
\begin{tabular}{c}
\includegraphics[angle=0,width=1.0\linewidth]{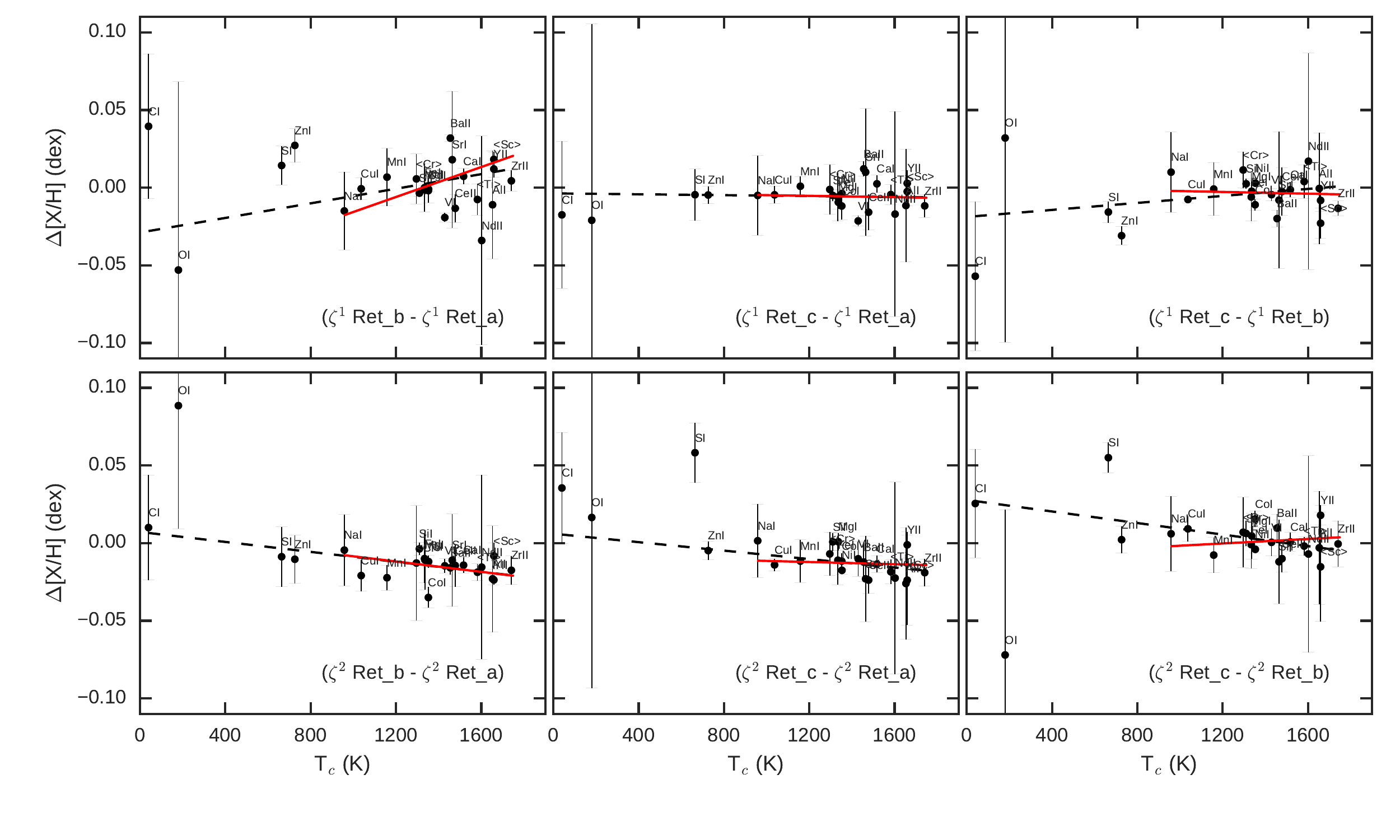}
\end{tabular}
\end{center}
\vspace{-1.0cm}
\caption{Differential abundances against condensation temperature for \zetone \ and \zettwo , derived from three highest S/N individual spectra.
The black dashed line and red line are the same as
in Fig.~\ref{fig_rel_sun}.}
\label{fig_zet1abc_zet2abs}
\end{figure*}

We also use a Bayesian approach to assess the presence of a dependence of [X/H] on \tc \, and to calculate
the slope of the dependence. 
It has been shown that p-value analysis, even if very widespread, is not as robust as expected; the assessment of the significance of a 
correlation is further complicated when the p-value is close to the significance level. 
In \citet{Figueira-16} we provided a straightforward alternative Bayesian approach to the assessment and characterization of a 
correlation in a dataset. However, relying on the Pearson correlation coefficient and Sperman's rank, \citet{Figueira-16} did not consider the 
impact of error bars on the measurements. We consider the case of a linear regression in which the value of each ordinate $Y_i$ 
follows a Gaussian distribution of center dictated by the evaluation of a linear slope on the abscissa $X_i$ and variance given by $\sigma^2$. 
Using the MCMC (Markov Chain Monte Carlo) algorithm implemented in PyMC \citep{Patil-10}, and assuming uninformative Gaussian distributions of center 0 and $\sigma$\,=\,1000 as priors 
for the slope parameters and intersect, we estimated the best fit to the data. 
In Table\,\ref{tab:slopes} we present the \tc  \ slopes and the associated 95\% credible intervals (highest posterior density; HPD) calculated by
applying this Bayesian approach.

The Bayesian and frequentist analyses differ in important ways. Since both approaches are addressing and solving exactly
the same problem, the slopes calculated by the two methods are almost identical. However, since the CI (in frequentist analysis) and the HPD (in Bayesian approach)
have different meanings, their values can be different and should be interpreted differently \citep[e.g.,][]{Morey-15}. 
A 95\% CI is an interval that in repeated sampling has a 0.95 probability of containing the true value of the parameter,
i.e, if a large number of samples are used,  the true value of the slope 
will fall within the CI in 95\% of the cases. A 95\% HPD is the interval that contains the true value of slope with a probability of 0.95 given the current sample (data). 
This said, it should not come as a surprise that a 95\% HPD is usually much smaller than the 95\% CI, which makes most of the correlations significant
(at 95\% level). 

In Table\,\ref{tab:slopes} we present the \tc  \ slopes that were derived for different pairs of stars and spectra. The slopes 
are derived by considering all of the elements and also by considering only refractory elements that have \tc \ $>$ 900 K.

To understand better how the calculated \tc \ slopes are sensitive to the abundance uncertainties, we also calculated the slopes without considering 
the errors. In the frequentist approach giving the same error (which means the same weight) to all of the elements or not considering errors yields the same 
results, since the calculated slope value is only  sensitive to the relative weights of the points. In the Bayesian approach, the absolute error value of each point is
important when evaluating the significance of the correlations, i.e., assuming 0.01 dex error for all of the elements produces 
a result that is very different if an error of 0.1 dex was assumed for all of the elements.

In the Appendix we present all of the \tc \ trend calculations discussed in the main text, 
without considering the errors for the individual abundances, i.e., applying an ordinary linear regression (OLS).

\section{Results} \label{sec:results}

In Fig.~\ref{fig_rel_sun} we show the dependence of differential [X/H] abundances of \zetone \ and \zettwo \ relative to the Sun on the corresponding \tc. The abundances
are derived using the combined spectra and spectra that were used in S16.  Table\,\ref{tab:slopes} and the corresponding plot shows 
that there is no trend with \tc \ for \zetone, \ while \zettwo \ shows a significant trend when the frequentist p-values are considered 
for null-hypothesis rejection testing. This trend, seen both from the combined and S16 spectra, is heavily driven 
by SI and ZnI (due to their small errors and hence large weight) and disappears when only refractory elements with \tc \ $> 900$ K 
are considered.  We note that S16 did not find a significant \tc \ trend for the two stars when compared with the Sun.
However, direct comparison of the slopes is not straightforward since we use different elements than S16 and, furthermore, those investigators corrected the abundances for the Galactic chemical evolution effect,
while we did not.

Since the differential abundances of the stars relative to the Sun can be affected by several factors, such as GCE, age, or the relative effect of NLTE 
on stars with similar, but not exactly the same stellar parameters, and we are interested in testing whether the two stars of the binary system show different 
refractory-to-volatile element ratios, in  Fig.~\ref{fig_rel_z1} we plot the differential abundances of \zetone \ relative to \zettwo. 
The plot and corresponding
Table\,\ref{tab:slopes} shows that there is a strong and significant correlation with the \tc \ in both cases when all of the elements 
and only refractory elements are considered for the linear regression. The slopes derived from the combined and S16 spectra 
are very similar, but the errors of the slopes for the case of combined spectra are smaller and the significance is higher. 
Following the interpretation of \citet{Melendez-09}, in S16 the authors 
based on this result proposed that refractory elements depleted in \zettwo \ are locked up in the debris disk that the star hosts.

To test the hypothesis that the observed \tc \ trend is due to the presence of a debris around \zettwo ,\ we performed the following test. We used
the three highest S/N 
individual spectra for each star (\zetone \ and \zettwo) and derived differential abundances for each star using different spectra as reference. The results 
are plotted in Fig.~\ref{fig_zet1abc_zet2abs} and presented in Table\,\ref{tab:slopes}. This test clearly shows that the slope 
of the \tc \ trend significantly  changes depending on which pair of spectra are compared. 
It is interesting to note that some of the observed trends are relatively significant (e.g., \zettwo\_b -- \zettwo\_a). 
These results, in turn, mean that if different spectra of the two stars 
are used to derive the \tc \ trend between the two stars, very different results can be obtained, ranging in significance from nonsignificant 
(i.e., having a slope compatible with zero) to very significant.

It is difficult to identify the main reason(s) of the observed \tc \ trend between different spectra of the same star. 
All of the spectra are observed with the same, and most stable spectrograph, HARPS, which minimizes the possible effects of spectral 
resolution and instrumental profile. While the three spectra of \zettwo \  
are taken at different dates (on 11 August 2006, 27 August 2009, and 31 January 2010, for \zettwo\_a, \zettwo\_b, and \zettwo\_c, respectively), the selected three 
individual high S/N spectra of \zetone \ are taken on the same night (15 November 2005). This means that time-dependent systematics 
are much less likely to play an important role for the observed trend for \zetone. This could be the reason that the \tc \ slopes
for this star are less significant compared to those of its companion. 
The two stars exhibit different levels of activity; 
\zetone \ ($<\log R_{HK}^{'}> \ = -4.662$) is more active than \zettwo \ ($<\log R_{HK}^{'}> \ = -4.892$) \citep{Zechmeister-13}.
Although \zetone \ is more active than its companion, the stellar variability should not play a significant role in the observed difference in the chemical abundances 
between the different spectra of the star since, again, the observations are carried out on the same night.

Recently, \citet{Bedell-14} analyzed solar spectra observed with different instruments, from different asteroids, and at different times, i.e, conditions.
The authors reached a conclusion that the major effect on differential relative abundances is caused by the use of different instruments. They also found no significant (more than 2$\sigma$) \tc \ trend between 
the different spectra they used.

One of the main factors that can produce (or affect) the \tc \ trend is the correct derivation of stellar parameters and their errors. 
Different elements (and corresponding spectral lines) show different sensitivities to atmospheric temperature 
(depending on the excitation potential), surface gravity (depending on the ionization state), and stellar metallicity 
\citep[see, e.g.,][]{Adibekyan-12c}. 

Another very important factor that can determine the \tc \ slope is the error estimation for the individual elements. Since we use 
the errors as weights for each abundance, the elements with the smallest errors have the largest weights and may determine the 
slope and its confidence interval. As discussed in Sect.\,\ref{sec:parameters}, the errors can be underestimated only if two
lines are used and the line-to-line dispersion is used to estimate the errors on the EWs. 
It is also important to mention the high weights of elements with a large number of lines in the derivation of the \tc \ slopes.
Since  the error of the abundances is 
calculated as a $\sigma /\sqrt{(n-1)}$, the elements such as Ni and Ti, which have many spectral lines, show the smallest errors
and, hence, have a key role in the determination of the slope value.

The importance of the errors of individual chemical abundances is well illustrated in  
Figs.~\ref{fig_rel_sun_without_errors},\ref{fig_rel_z1_without_errors},\ref{fig_zet1abc_zet2abs_without_errors}, and 
in Table\,\ref{tab:slopes_without_errors}. In these plots and table, we present the derivation of the \tc \ slopes for the
above discussed stars and spectra,  but applying OLS,  i.e., no errors (or equal errors) are considered. We can clearly see that for the same star or spectra the OLS and WLS give very different results. The Table\,\ref{tab:slopes_without_errors}
also shows that the most significant trends are observed for (\zetone\_comb -- \zettwo\_comb), (\zettwo\_b -- \zettwo\_a), and 
(\zettwo\_c -- \zettwo\_a). Moreover, the only significant trend that remains when considering only the refractory elements
with \tc \ $>$ 900K is for the (\zettwo\_c -- \zettwo\_a) pair.

\section{Conclusion} \label{sec:conclusion}

The $\zeta$ Reticuli binary system, composed of two solar analogs, is a very interesting and well-known system not only in the scientific literature, 
but also in science fiction literature\footnote{See \url{https://en.wikipedia.org/wiki/Stars\_and\_planetary\_systems\_in\_fiction\#Zeta\_Reticuli}} %
and movies,\footnote{\url{https://en.wikipedia.org/wiki/The_UFO_Incident}} %
as a word of  Zeta Reticulans\footnote{\url{https://en.wikipedia.org/wiki/Grey_alien}}%
.

Both stars of this binary system are in planet search programs and the presence of short-period small (or massive) planets or 
larger period massive planets can be excluded (see S16). However, there is indirect evidence that the \zettwo \ may have a massive eccentric 
companion that perturbed its eccentric debris disk \citep{Faramaz-14}. 

We used several high-quality (S/N $>$ 360) individual and combined spectra (S/N of 1300 and 3000) of \zetone \ and \zettwo \ to revisit the results 
obtained in S16, namely that \zettwo \ shows a deficit of refractory elements (relative to volatiles) when compared to \zetone \ probably because of the 
presence of the debris disk that \zettwo \ hosts. We first derived the stellar parameters using the classical (nondifferential) method, 
then applied careful line-by-line differential abundance analysis of the stars relative to the Sun and relative to each other.

When we consider the combined, highest S/N spectra of the two stars we obtain similar \tc \ trend as was reported in S16.
The trend exists when all of the elements and only refractory elements are considered in the derivation of the \tc \ slope. 
We also show that when comparing the chemical abundances of the same individual stars (\zetone \ with \zetone \ and   \zettwo \ with \zettwo)
derived from different individual spectra, we observe different and sometimes significant \tc \ trends. The trends observed between different individual
spectra of \zetone \ that are observed during the same night, are less significant than those observed for the \zettwo \ that
are observed on different nights.

In this context, there is no consensus on the results of the \tc \ trend
for some individual systems \citep[e.g., 16 Cyg AB;][]{Laws-01, Takeda-05, Schuler-11a, TucciMaia-14}. The reported differences
can be due to the same sort of effect that we discussed in this study.

Our results show that when studying  tiny chemical abundance trends with condensation temperature, it is very important to use very high-quality combined spectra.
The combination of several spectra increase the S/N and may minimize possible time-dependent effects.
The results also show that there are other nonastrophysical factors (such as over- or underestimation of the errors of the individual elements)
that may be responsible for the observed \tc \ trends and indicate that 
one should be very careful when analyzing very subtle differences in chemical abundances between stars. 
However, we should stress that this result does not imply that all of the observed \tc \ trends do not have an astrophysical origin.

\begin{acknowledgements}

{This work was supported by Funda\c{c}\~ao para a Ci\^encia e Tecnologia (FCT) through national funds (project ref. PTDC/FIS-AST/7073/2014) 
and by FEDER through COMPETE2020 (project ref. POCI-01-0145-FEDER-007672). This work was also supported by FCT through the research grants (ref. PTDC/FIS-AST/7073/2014
and ref. PTDC/FIS-AST/1526/2014) through national funds and by FEDER through COMPETE2020 (ref. POCI-01-0145-FEDER-016880 and ref. POCI-01-0145-FEDER-016886).
This work results within the collaboration of the COST Action  TD1308.
P.F., N.C.S., and S.G.S. also acknowledge the support from FCT through Investigador FCT contracts of reference IF/01037/2013, IF/00169/2012, 
and IF/00028/2014, respectively, and POPH/FSE (EC) by FEDER funding through the program ``Programa Operacional de Factores de Competitividade - COMPETE''. 
PF further acknowledges support from FCT in the form of an exploratory project of reference IF/01037/2013CP1191/CT0001. 
V.A. and E.D.M acknowledge the support from the FCT in the form of the grants SFRH/BPD/70574/2010 and SFRH/BPD/76606/2011, respectively.
V.A also acknowledges the support from COST Action TD1308 through STSM grant with reference Number: COST-STSM-TD1308-32051.
G.I. acknowledges financial support from the Spanish Ministry project MINECO AYA2011-29060.
JIGH acknowledges financial support from the Spanish Ministry of Economy and Competitiveness (MINECO) 
under the 2013 Ram\'{o}n y Cajal program MINECO RYC-2013-14875, and the Spanish ministry project MINECO AYA2014-56359-P.}

\end{acknowledgements}

\bibliography{refbib}

\begin{thebibliography}{63}
\expandafter\ifx\csname natexlab\endcsname\relax\def\natexlab#1{#1}\fi

\bibitem[{{Adibekyan} {et~al.}(2015{\natexlab{a}}){Adibekyan}, {Figueira}, \&
  {Santos}}]{Adibekyan-16}
{Adibekyan}, V., {Figueira}, P., \& {Santos}, N.~C. 2015{\natexlab{a}},
  [arXiv:1509.02429]

\bibitem[{{Adibekyan} {et~al.}(2015{\natexlab{b}}){Adibekyan}, {Figueira},
  {Santos}, {Sousa}, {Faria}, {Delgado-Mena}, {Oshagh}, {Tsantaki}, {Hakobyan},
  {Gonz{\'a}lez Hern{\'a}ndez}, {Su{\'a}rez-Andr{\'e}s}, \&
  {Israelian}}]{Adibekyan-15b}
{Adibekyan}, V., {Figueira}, P., {Santos}, N.~C., {et~al.} 2015{\natexlab{b}},
  \aap, 583, A94

\bibitem[{{Adibekyan} {et~al.}(2015{\natexlab{c}}){Adibekyan}, {Santos},
  {Figueira}, {Dorn}, {Sousa}, {Delgado-Mena}, {Israelian}, {Hakobyan}, \&
  {Mordasini}}]{Adibekyan-15a}
{Adibekyan}, V., {Santos}, N.~C., {Figueira}, P., {et~al.} 2015{\natexlab{c}},
  \aap, 581, L2

\bibitem[{{Adibekyan} {et~al.}(2012{\natexlab{a}}){Adibekyan}, {Delgado Mena},
  {Sousa}, {Santos}, {Israelian}, {Gonz{\'a}lez Hern{\'a}ndez}, {Mayor}, \&
  {Hakobyan}}]{Adibekyan-12a}
{Adibekyan}, V.~Z., {Delgado Mena}, E., {Sousa}, S.~G., {et~al.}
  2012{\natexlab{a}}, \aap, 547, A36

\bibitem[{{Adibekyan} {et~al.}(2014){Adibekyan}, {Gonz{\'a}lez Hern{\'a}ndez},
  {Delgado Mena}, {Sousa}, {Santos}, {Israelian}, {Figueira}, \& {Bertran de
  Lis}}]{Adibekyan-14}
{Adibekyan}, V.~Z., {Gonz{\'a}lez Hern{\'a}ndez}, J.~I., {Delgado Mena}, E.,
  {et~al.} 2014, \aap, 564, L15

\bibitem[{{Adibekyan} {et~al.}(2012{\natexlab{b}}){Adibekyan}, {Santos},
  {Sousa}, {Israelian}, {Delgado Mena}, {Gonz{\'a}lez Hern{\'a}ndez}, {Mayor},
  {Lovis}, \& {Udry}}]{Adibekyan-12b}
{Adibekyan}, V.~Z., {Santos}, N.~C., {Sousa}, S.~G., {et~al.}
  2012{\natexlab{b}}, \aap, 543, A89

\bibitem[{{Adibekyan} {et~al.}(2012{\natexlab{c}}){Adibekyan}, {Sousa},
  {Santos}, {Delgado Mena}, {Gonz{\'a}lez Hern{\'a}ndez}, {Israelian}, {Mayor},
  \& {Khachatryan}}]{Adibekyan-12c}
{Adibekyan}, V.~Z., {Sousa}, S.~G., {Santos}, N.~C., {et~al.}
  2012{\natexlab{c}}, \aap, 545, A32

\bibitem[{{Asplund} {et~al.}(2009){Asplund}, {Grevesse}, {Sauval}, \&
  {Scott}}]{Asplund-09}
{Asplund}, M., {Grevesse}, N., {Sauval}, A.~J., \& {Scott}, P. 2009, \araa, 47,
  481

\bibitem[{{Bedell} {et~al.}(2014){Bedell}, {Mel{\'e}ndez}, {Bean},
  {Ram{\'{\i}}rez}, {Leite}, \& {Asplund}}]{Bedell-14}
{Bedell}, M., {Mel{\'e}ndez}, J., {Bean}, J.~L., {et~al.} 2014, \apj, 795, 23

\bibitem[{{Bertran de Lis} {et~al.}(2015){Bertran de Lis}, {Delgado Mena},
  {Adibekyan}, {Santos}, \& {Sousa}}]{Bertrandelis-15}
{Bertran de Lis}, S., {Delgado Mena}, E., {Adibekyan}, V.~Z., {Santos}, N.~C.,
  \& {Sousa}, S.~G. 2015, \aap, 576, A89

\bibitem[{{Biazzo} {et~al.}(2015){Biazzo}, {Gratton}, {Desidera}, {Lucatello},
  {Sozzetti}, {Bonomo}, {Damasso}, {Gandolfi}, {Affer}, {Boccato}, {Borsa},
  {Claudi}, {Cosentino}, {Covino}, {Knapic}, {Lanza}, {Maldonado}, {Marzari},
  {Micela}, {Molaro}, {Pagano}, {Pedani}, {Pillitteri}, {Piotto}, {Poretti},
  {Rainer}, {Santos}, {Scandariato}, \& {Zanmar Sanchez}}]{Biazzo-15}
{Biazzo}, K., {Gratton}, R., {Desidera}, S., {et~al.} 2015, \aap, 583, A135

\bibitem[{{Bond} {et~al.}(2010){Bond}, {O'Brien}, \& {Lauretta}}]{Bond-10}
{Bond}, J.~C., {O'Brien}, D.~P., \& {Lauretta}, D.~S. 2010, \apj, 715, 1050

\bibitem[{{Cayrel}(1988)}]{Cayrel-98}
{Cayrel}, R. 1988, in IAU Symposium, Vol. 132, The Impact of Very High S/N
  Spectroscopy on Stellar Physics, ed. G.~{Cayrel de Strobel} \& M.~{Spite},
  345

\bibitem[{{Delgado Mena} {et~al.}(2015){Delgado Mena}, {Bertr{\'a}n de Lis},
  {Adibekyan}, {Sousa}, {Figueira}, {Mortier}, {Gonz{\'a}lez Hern{\'a}ndez},
  {Tsantaki}, {Israelian}, \& {Santos}}]{Delgado-15}
{Delgado Mena}, E., {Bertr{\'a}n de Lis}, S., {Adibekyan}, V.~Z., {et~al.}
  2015, \aap, 576, A69

\bibitem[{{Delgado Mena} {et~al.}(2010){Delgado Mena}, {Israelian},
  {Gonz{\'a}lez Hern{\'a}ndez}, {Bond}, {Santos}, {Udry}, \&
  {Mayor}}]{Delgado-10}
{Delgado Mena}, E., {Israelian}, G., {Gonz{\'a}lez Hern{\'a}ndez}, J.~I.,
  {et~al.} 2010, \apj, 725, 2349

\bibitem[{{Delgado Mena} {et~al.}(2014){Delgado Mena}, {Israelian},
  {Gonz{\'a}lez Hern{\'a}ndez}, {Sousa}, {Mortier}, {Santos}, {Adibekyan},
  {Fernandes}, {Rebolo}, {Udry}, \& {Mayor}}]{Delgado-14}
{Delgado Mena}, E., {Israelian}, G., {Gonz{\'a}lez Hern{\'a}ndez}, J.~I.,
  {et~al.} 2014, \aap, 562, A92

\bibitem[{{Dorn} {et~al.}(2015){Dorn}, {Khan}, {Heng}, {Connolly}, {Alibert},
  {Benz}, \& {Tackley}}]{Dorn-15}
{Dorn}, C., {Khan}, A., {Heng}, K., {et~al.} 2015, \aap, 577, A83

\bibitem[{{Ecuvillon} {et~al.}(2006){Ecuvillon}, {Israelian}, {Santos},
  {Mayor}, \& {Gilli}}]{Ecuvillon-06}
{Ecuvillon}, A., {Israelian}, G., {Santos}, N.~C., {Mayor}, M., \& {Gilli}, G.
  2006, \aap, 449, 809

\bibitem[{{Faramaz} {et~al.}(2014){Faramaz}, {Beust}, {Th{\'e}bault},
  {Augereau}, {Bonsor}, {del Burgo}, {Ertel}, {Marshall}, {Milli},
  {Montesinos}, {Mora}, {Bryden}, {Danchi}, {Eiroa}, {White}, \&
  {Wolf}}]{Faramaz-14}
{Faramaz}, V., {Beust}, H., {Th{\'e}bault}, P., {et~al.} 2014, \aap, 563, A72

\bibitem[{{Figueira} {et~al.}(2016){Figueira}, {Faria}, {Adibekyan}, {Oshagh},
  \& {Santos}}]{Figueira-16}
{Figueira}, P., {Faria}, J.~P., {Adibekyan}, V.~Z., {Oshagh}, M., \& {Santos},
  N.~C. 2016, [arXiv:1601.05107]

\bibitem[{{Gaidos}(2015)}]{Gaidos-15}
{Gaidos}, E. 2015, \apj, 804, 40

\bibitem[{{Gilli} {et~al.}(2006){Gilli}, {Israelian}, {Ecuvillon}, {Santos}, \&
  {Mayor}}]{Gilli-06}
{Gilli}, G., {Israelian}, G., {Ecuvillon}, A., {Santos}, N.~C., \& {Mayor}, M.
  2006, \aap, 449, 723

\bibitem[{{Gonzalez}(1997)}]{Gonzalez-97}
{Gonzalez}, G. 1997, \mnras, 285, 403

\bibitem[{{Gonz{\'a}lez Hern{\'a}ndez} {et~al.}(2013){Gonz{\'a}lez
  Hern{\'a}ndez}, {Delgado-Mena}, {Sousa}, {Israelian}, {Santos}, {Adibekyan},
  \& {Udry}}]{GH-13}
{Gonz{\'a}lez Hern{\'a}ndez}, J.~I., {Delgado-Mena}, E., {Sousa}, S.~G.,
  {et~al.} 2013, \aap, 552, A6

\bibitem[{{Gonz{\'a}lez Hern{\'a}ndez} {et~al.}(2010){Gonz{\'a}lez
  Hern{\'a}ndez}, {Israelian}, {Santos}, {Sousa}, {Delgado-Mena}, {Neves}, \&
  {Udry}}]{GH-10}
{Gonz{\'a}lez Hern{\'a}ndez}, J.~I., {Israelian}, G., {Santos}, N.~C., {et~al.}
  2010, \apj, 720, 1592

\bibitem[{{Kang} {et~al.}(2011){Kang}, {Lee}, \& {Kim}}]{Kang-11}
{Kang}, W., {Lee}, S.-G., \& {Kim}, K.-M. 2011, \apj, 736, 87

\bibitem[{{Kurucz}(1993)}]{Kurucz-93}
{Kurucz}, R.~L. 1993, {SYNTHE spectrum synthesis programs and line data}

\bibitem[{{Laws} \& {Gonzalez}(2001)}]{Laws-01}
{Laws}, C. \& {Gonzalez}, G. 2001, \apj, 553, 405

\bibitem[{{Liu} {et~al.}(2014){Liu}, {Asplund}, {Ram{\'{\i}}rez}, {Yong}, \&
  {Mel{\'e}ndez}}]{Liu-14}
{Liu}, F., {Asplund}, M., {Ram{\'{\i}}rez}, I., {Yong}, D., \& {Mel{\'e}ndez},
  J. 2014, \mnras, 442, L51

\bibitem[{{Lodders}(2003)}]{Lodders-03}
{Lodders}, K. 2003, \apj, 591, 1220

\bibitem[{{Mack} {et~al.}(2016){Mack}, {Stassun}, {Schuler}, {Hebb}, \&
  {Pepper}}]{Mack-16}
{Mack}, III, C.~E., {Stassun}, K.~G., {Schuler}, S.~C., {Hebb}, L., \&
  {Pepper}, J.~A. 2016, [arXiv:1601.00018]

\bibitem[{{Maldonado} {et~al.}(2015){Maldonado}, {Eiroa}, {Villaver},
  {Montesinos}, \& {Mora}}]{Maldonado-15}
{Maldonado}, J., {Eiroa}, C., {Villaver}, E., {Montesinos}, B., \& {Mora}, A.
  2015, \aap, 579, A20

\bibitem[{{Maldonado} \& {Villaver}(2016)}]{Maldonado-16}
{Maldonado}, J. \& {Villaver}, E. 2016, \aap, 588, A98

\bibitem[{{Mayor} {et~al.}(2003){Mayor}, {Pepe}, {Queloz}, {Bouchy},
  {Rupprecht}, {Lo Curto}, {Avila}, {Benz}, {Bertaux}, {Bonfils}, {Dall},
  {Dekker}, {Delabre}, {Eckert}, {Fleury}, {Gilliotte}, {Gojak}, {Guzman},
  {Kohler}, {Lizon}, {Longinotti}, {Lovis}, {Megevand}, {Pasquini}, {Reyes},
  {Sivan}, {Sosnowska}, {Soto}, {Udry}, {van Kesteren}, {Weber}, \&
  {Weilenmann}}]{Mayor-03}
{Mayor}, M., {Pepe}, F., {Queloz}, D., {et~al.} 2003, The Messenger, 114, 20

\bibitem[{{Mel{\'e}ndez} {et~al.}(2009){Mel{\'e}ndez}, {Asplund}, {Gustafsson},
  \& {Yong}}]{Melendez-09}
{Mel{\'e}ndez}, J., {Asplund}, M., {Gustafsson}, B., \& {Yong}, D. 2009, \apjl,
  704, L66

\bibitem[{Morey {et~al.}(2015)Morey, Hoekstra, Rouder, Lee, \&
  Wagenmakers}]{Morey-15}
Morey, R.~D., Hoekstra, R., Rouder, J.~N., Lee, M.~D., \& Wagenmakers, E.-J.
  2015, Psychonomic Bulletin \& Review, 23, 103

\bibitem[{{Neuforge-Verheecke} \& {Magain}(1997)}]{Neuforge-97}
{Neuforge-Verheecke}, C. \& {Magain}, P. 1997, \aap, 328, 261

\bibitem[{{Nissen}(2015)}]{Nissen-15}
{Nissen}, P.~E. 2015, \aap, 579, A52

\bibitem[{{{\"O}nehag} {et~al.}(2014){{\"O}nehag}, {Gustafsson}, \&
  {Korn}}]{Onehag-14}
{{\"O}nehag}, A., {Gustafsson}, B., \& {Korn}, A. 2014, \aap, 562, A102

\bibitem[{Patil {et~al.}(2010)Patil, Huard, \& Fonnesbeck}]{Patil-10}
Patil, A., Huard, D., \& Fonnesbeck, C.~J. 2010, Journal of Statistical
  Software, 35, 1

\bibitem[{{Prochaska} {et~al.}(2000){Prochaska}, {Naumov}, {Carney},
  {McWilliam}, \& {Wolfe}}]{Prochaska-00}
{Prochaska}, J.~X., {Naumov}, S.~O., {Carney}, B.~W., {McWilliam}, A., \&
  {Wolfe}, A.~M. 2000, \aj, 120, 2513

\bibitem[{{Ram{\'{\i}}rez} {et~al.}(2015){Ram{\'{\i}}rez}, {Khanal}, {Aleo},
  {Sobotka}, {Liu}, {Casagrande}, {Mel{\'e}ndez}, {Yong}, {Lambert}, \&
  {Asplund}}]{Ramirez-15}
{Ram{\'{\i}}rez}, I., {Khanal}, S., {Aleo}, P., {et~al.} 2015, \apj, 808, 13

\bibitem[{{Ram{\'{\i}}rez} {et~al.}(2009){Ram{\'{\i}}rez}, {Mel{\'e}ndez}, \&
  {Asplund}}]{Ramirez-09}
{Ram{\'{\i}}rez}, I., {Mel{\'e}ndez}, J., \& {Asplund}, M. 2009, \aap, 508, L17

\bibitem[{{Ram{\'{\i}}rez} {et~al.}(2014){Ram{\'{\i}}rez}, {Mel{\'e}ndez}, \&
  {Asplund}}]{Ramirez-14}
{Ram{\'{\i}}rez}, I., {Mel{\'e}ndez}, J., \& {Asplund}, M. 2014, \aap, 561, A7

\bibitem[{{Reddy} {et~al.}(2003){Reddy}, {Tomkin}, {Lambert}, \& {Allende
  Prieto}}]{Reddy-03}
{Reddy}, B.~E., {Tomkin}, J., {Lambert}, D.~L., \& {Allende Prieto}, C. 2003,
  \mnras, 340, 304

\bibitem[{{Robinson} {et~al.}(2006){Robinson}, {Laughlin}, {Bodenheimer}, \&
  {Fischer}}]{Robinson-06}
{Robinson}, S.~E., {Laughlin}, G., {Bodenheimer}, P., \& {Fischer}, D. 2006,
  \apj, 643, 484

\bibitem[{{Saffe} {et~al.}(2015){Saffe}, {Flores}, \& {Buccino}}]{Saffe-15}
{Saffe}, C., {Flores}, M., \& {Buccino}, A. 2015, \aap, 582, A17

\bibitem[{{Saffe} {et~al.}(2016){Saffe}, {Flores}, {Jaque Arancibia},
  {Buccino}, \& {Jofre}}]{Saffe-16}
{Saffe}, C., {Flores}, M., {Jaque Arancibia}, M., {Buccino}, A., \& {Jofre}, E.
  2016, [arXiv:1602.01320]

\bibitem[{{Santos} {et~al.}(2015){Santos}, {Adibekyan}, {Mordasini}, {Benz},
  {Delgado-Mena}, {Dorn}, {Buchhave}, {Figueira}, {Mortier}, {Pepe},
  {Santerne}, {Sousa}, \& {Udry}}]{Santos-15}
{Santos}, N.~C., {Adibekyan}, V., {Mordasini}, C., {et~al.} 2015, \aap, 580,
  L13

\bibitem[{{Schuler} {et~al.}(2011{\natexlab{a}}){Schuler}, {Cunha}, {Smith},
  {Ghezzi}, {King}, {Deliyannis}, \& {Boesgaard}}]{Schuler-11a}
{Schuler}, S.~C., {Cunha}, K., {Smith}, V.~V., {et~al.} 2011{\natexlab{a}},
  \apjl, 737, L32

\bibitem[{{Schuler} {et~al.}(2011{\natexlab{b}}){Schuler}, {Flateau}, {Cunha},
  {King}, {Ghezzi}, \& {Smith}}]{Schuler-11}
{Schuler}, S.~C., {Flateau}, D., {Cunha}, K., {et~al.} 2011{\natexlab{b}},
  \apj, 732, 55

\bibitem[{{Smith} {et~al.}(2001){Smith}, {Cunha}, \& {Lazzaro}}]{Smith-01}
{Smith}, V.~V., {Cunha}, K., \& {Lazzaro}, D. 2001, \aj, 121, 3207

\bibitem[{{Sneden}(1973)}]{Sneden-73}
{Sneden}, C.~A. 1973, PhD thesis, THE UNIVERSITY OF TEXAS AT AUSTIN.

\bibitem[{{Sousa}(2014)}]{Sousa-14}
{Sousa}, S.~G. 2014, [arXiv:1407.5817]

\bibitem[{{Sousa} {et~al.}(2015){Sousa}, {Santos}, {Adibekyan}, {Delgado-Mena},
  \& {Israelian}}]{Sousa-15}
{Sousa}, S.~G., {Santos}, N.~C., {Adibekyan}, V., {Delgado-Mena}, E., \&
  {Israelian}, G. 2015, \aap, 577, A67

\bibitem[{{Sozzetti} {et~al.}(2006){Sozzetti}, {Yong}, {Carney}, {Laird},
  {Latham}, \& {Torres}}]{Sozzetti-06}
{Sozzetti}, A., {Yong}, D., {Carney}, B.~W., {et~al.} 2006, \aj, 131, 2274

\bibitem[{{Takeda}(2005)}]{Takeda-05}
{Takeda}, Y. 2005, \pasj, 57, 83

\bibitem[{{Takeda} {et~al.}(2001){Takeda}, {Sato}, {Kambe}, {Aoki}, {Honda},
  {Kawanomoto}, {Masuda}, {Izumiura}, {Watanabe}, {Koyano}, {Maehara},
  {Norimoto}, {Okuda}, {Shimizu}, {Uraguchi}, {Yanagisawa}, {Yoshida},
  {Miyama}, \& {Ando}}]{Takeda-01}
{Takeda}, Y., {Sato}, B., {Kambe}, E., {et~al.} 2001, \pasj, 53, 1211

\bibitem[{{Teske} {et~al.}(2015){Teske}, {Ghezzi}, {Cunha}, {Smith}, {Schuler},
  \& {Bergemann}}]{Teske-15}
{Teske}, J.~K., {Ghezzi}, L., {Cunha}, K., {et~al.} 2015, \apjl, 801, L10

\bibitem[{{Teske} {et~al.}(2016){Teske}, {Khanal}, \&
  {Ram{\'{\i}}rez}}]{Teske-16}
{Teske}, J.~K., {Khanal}, S., \& {Ram{\'{\i}}rez}, I. 2016, [arXiv:1601.01731]

\bibitem[{{Thiabaud} {et~al.}(2014){Thiabaud}, {Marboeuf}, {Alibert}, {Cabral},
  {Leya}, \& {Mezger}}]{Thiabaud-14}
{Thiabaud}, A., {Marboeuf}, U., {Alibert}, Y., {et~al.} 2014, \aap, 562, A27

\bibitem[{{Tucci Maia} {et~al.}(2014){Tucci Maia}, {Mel{\'e}ndez}, \&
  {Ram{\'{\i}}rez}}]{TucciMaia-14}
{Tucci Maia}, M., {Mel{\'e}ndez}, J., \& {Ram{\'{\i}}rez}, I. 2014, \apjl, 790,
  L25

\bibitem[{{Zechmeister} {et~al.}(2013){Zechmeister}, {K{\"u}rster}, {Endl}, {Lo
  Curto}, {Hartman}, {Nilsson}, {Henning}, {Hatzes}, \&
  {Cochran}}]{Zechmeister-13}
{Zechmeister}, M., {K{\"u}rster}, M., {Endl}, M., {et~al.} 2013, \aap, 552, A78

\end{thebibliography}

\Online

\begin{appendix}

\section{\tc \ slope in OLS}

In this section we present the calculation of the \tc \ slopes for the stars without considering the errors of the individual
chemical abundances.

\begin{figure*}
\begin{center}
\begin{tabular}{c}
\includegraphics[angle=0,width=1.0\linewidth]{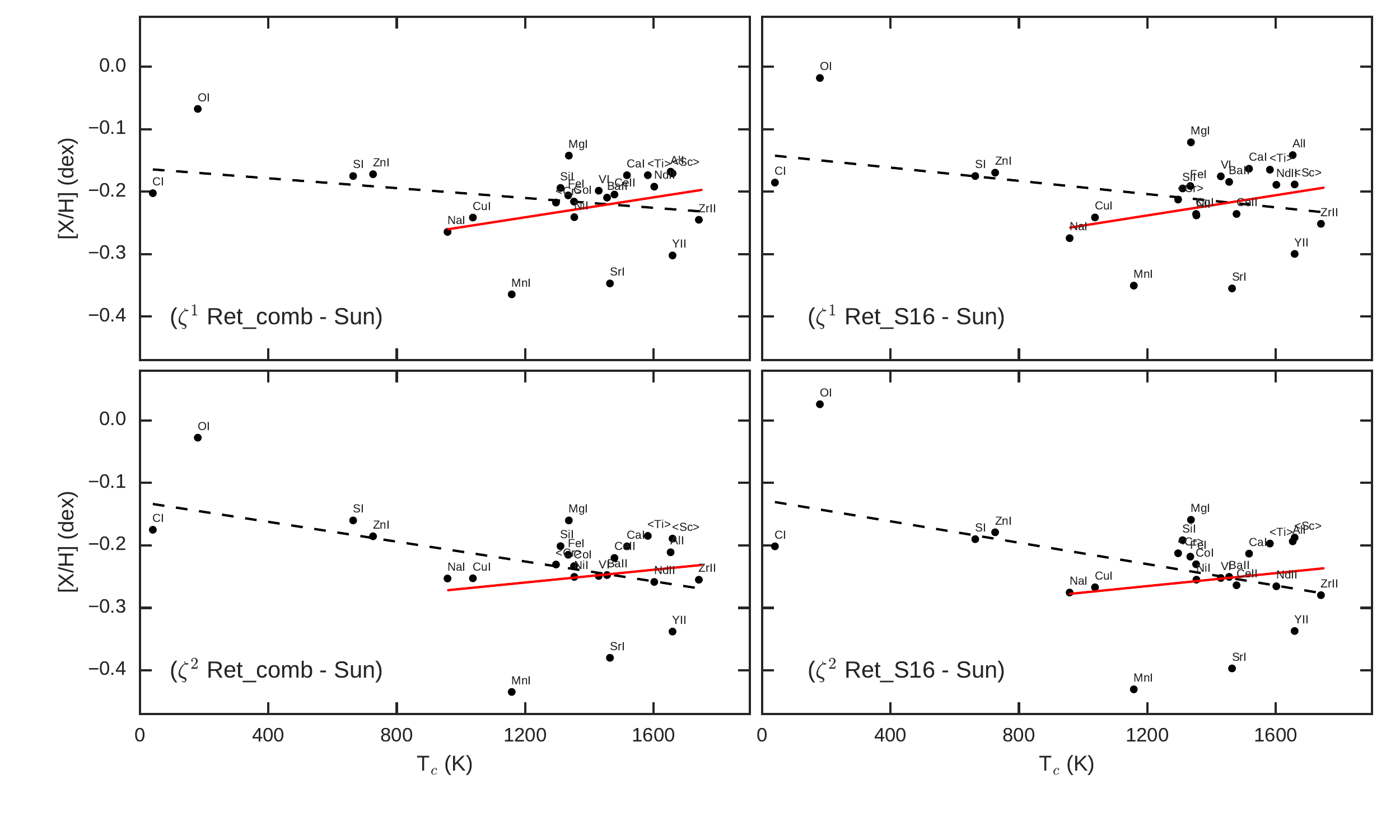}
\end{tabular}
\end{center}
\vspace{-1.0cm}
\caption{Same as Fig.~\ref{fig_rel_sun}, but the slopes are calculated without considering the errors of the individual abundances.}
\label{fig_rel_sun_without_errors}
\end{figure*}

\begin{figure*}
\begin{center}
\begin{tabular}{c}
\includegraphics[angle=0,width=0.6\linewidth]{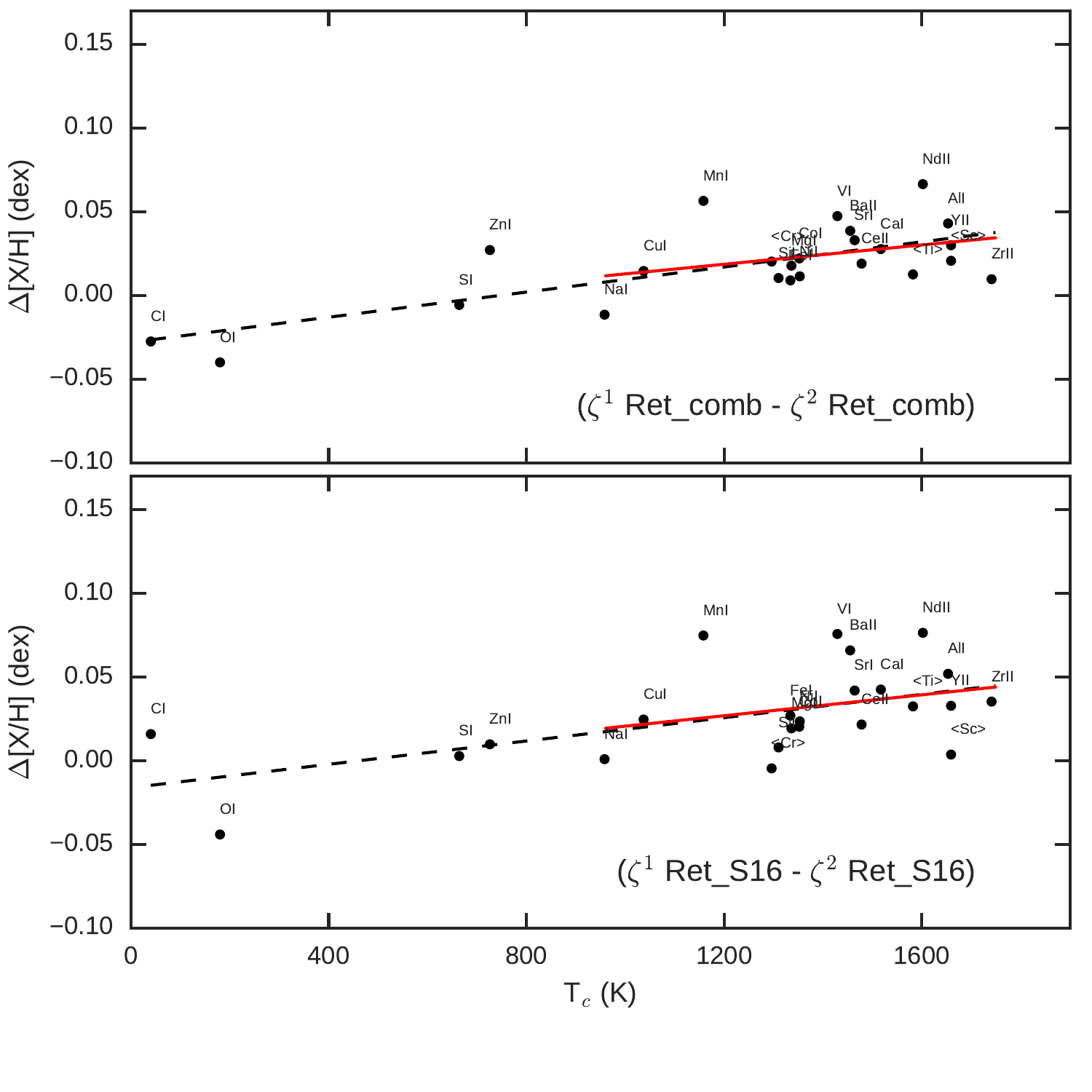}
\end{tabular}
\end{center}
\vspace{-1.cm}
\caption{Same as Fig.~\ref{fig_rel_z1}, but the slopes are calculated without considering the errors of the individual abundances.}
\label{fig_rel_z1_without_errors}
\end{figure*}

\begin{table*}[t!]

\setlength{\tabcolsep}{3pt}
\caption{\label{tab:slopes_without_errors} Slopes of the [X/H] vs. \tc \ for different pairs of spectra. A frequentist approach is chosen to derive 
the slopes and their uncertainties without considering the errors on individual element abundances.}
\tiny
\begin{center}
\begin{tabular}{lccc|ccc}

\hline\hline
Star -- Reference &\multicolumn{3}{c}{OLS (all elements)}  & \multicolumn{3}{c}{OLS (refractories)} \\ \cmidrule(l){2-4}\cmidrule(l){5-7}
 & \multicolumn{1}{c}{Slope$\pm \sigma$} & 95\% CI &  P(F-stat) & \multicolumn{1}{c}{Slope$\pm \sigma$} & 95\% CI &  P(F-stat) \\
\hline
\zetone\_comb -- Sun  & -3.94$\pm$2.88 & [-9.91, 2.04] &0.186& 7.97$\pm$ 6.33 & [-5.33, 21.28] &0.224 \tabularnewline
\zetone\_S16 -- Sun  & -5.30$\pm$3.21 & [-11.97, 1.36] &0.113& 8.09$\pm$ 6.88 & [-6.36, 22.54] &0.255 \tabularnewline
\zettwo\_comb -- Sun  & -7.94$\pm$3.32 & [-14.83, -1.05] &0.026& 5.11$\pm$ 7.41 & [-10.46, 20.68] &0.499 \tabularnewline
\zettwo\_S16 -- Sun  & -8.57$\pm$3.67 & [-16.18, -0.96] &0.029& 5.18$\pm$ 7.66 & [-10.91, 21.27] &0.507 \tabularnewline
\zetone\_comb -- \zettwo\_comb  & 3.75$\pm$0.83 & [2.04, 5.47] & 1.6$\times10^{-4}$ & 2.88$\pm$ 1.97 & [-1.25, 7.01] &0.160 \tabularnewline
\zetone\_S16 -- \zettwo\_S16  & 3.48$\pm$1.13 & [1.14, 5.83] &0.005& 3.12$\pm$ 2.70 & [-2.56, 8.80] &0.263 \tabularnewline
\zetone\_b -- \zetone\_a  & -0.19$\pm$0.96 & [-2.18, 1.80] &0.846& 0.69$\pm$ 1.65 & [-2.77, 4.16] &0.679 \tabularnewline
\zetone\_c -- \zetone\_a  & 0.60$\pm$0.39 & [-0.21, 1.42] &0.140& -0.42$\pm$ 0.96 & [-2.43, 1.59] &0.669 \tabularnewline
\zetone\_c -- \zetone\_b  & 0.83$\pm$0.80 & [-0.82, 2.48] &0.310& -1.38$\pm$ 1.06 & [-3.62, 0.85] &0.210 \tabularnewline
\zettwo\_b -- \zettwo\_a  & -3.33$\pm$0.80 & [-5.00, -1.66] & 4.2$\times10^{-4}$ & -0.65$\pm$ 0.81 & [-2.35, 1.04] &0.430 \tabularnewline
\zettwo\_c -- \zettwo\_a  & -3.36$\pm$0.60 & [-4.59, -2.12] & 1.2$\times10^{-5}$ & -2.23$\pm$ 0.84 & [-4.00, -0.46] &0.016 \tabularnewline
\zettwo\_c -- \zettwo\_b  & 0.13$\pm$1.01 & [-1.96, 2.22] &0.899& -1.14$\pm$ 0.96 & [-3.16, 0.89] &0.254 \tabularnewline
\hline
\end{tabular}
\end{center}
Note: Units of the slopes are in 10$^{-5}$ dex K$^{-1}$.
\end{table*}

\begin{figure*}
\begin{center}
\begin{tabular}{c}
\includegraphics[angle=0,width=1.0\linewidth]{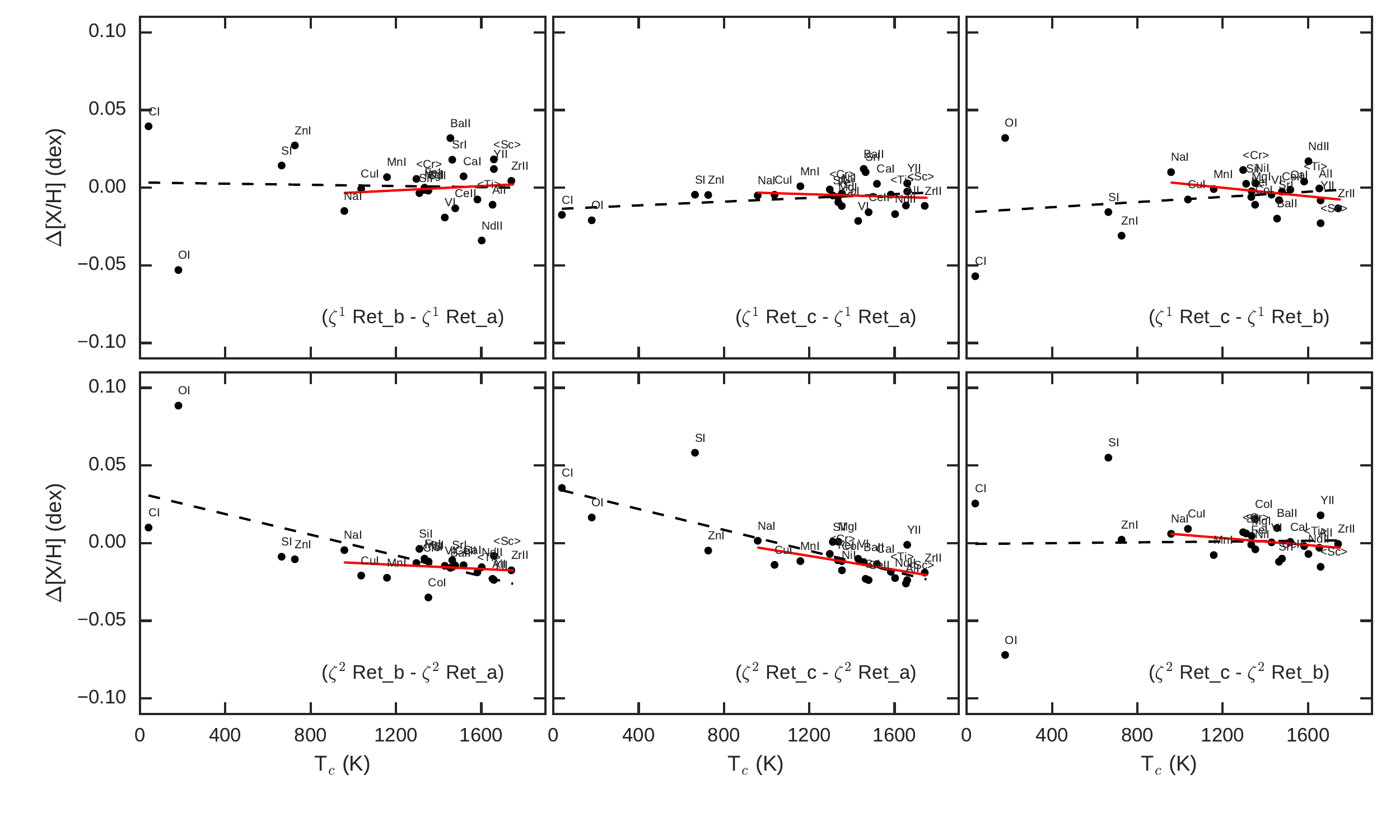}
\end{tabular}
\end{center}
\vspace{-1.0cm}
\caption{Same as Fig.~\ref{fig_zet1abc_zet2abs}, but the slopes are calculated without considering the errors of the individual abundances.}
\label{fig_zet1abc_zet2abs_without_errors}
\end{figure*}

\end{appendix}

\end{document}